\documentclass[aps,reprint,amsmath,amssymb,eqsecnumaps,pra]{revtex4-2}
\usepackage{graphicx}
\usepackage{dcolumn}
\usepackage{bm}
\usepackage{float}
\usepackage{xcolor}
\usepackage{lineno}
\usepackage[normalem]{ulem}

\newcommand{\sq}[1]{\sqrt{\smash[b]{#1}}}
\newcommand{\vta}{\vartheta}
\newcommand{\p}{\partial}
\newcommand{\ep}{\varepsilon}
\newcommand{\f}{\text{f}}

\newcommand{\s}{\text{s}}

\newcommand{\vD}{\varDelta}

\newcommand{\om}{\omega}
\newcommand{\nn}{\nonumber}
\newcommand{\ta}{\theta}

\newcommand{\cH}{{\cal H}}
\newcommand{\cF}{{\cal F}}

\newcommand{\cW}{{\cal W}}
\newcommand{\cN}{{\cal N}}

\newcommand{\ket}[1]{\vert{#1}\rangle}
\newcommand{\bra}[1]{\langle{#1}\vert}

\newcommand{\wh}{\widehat}
\newcommand{\wt}{\widetilde}
\newcommand{\be}{\begin{equation}}                                       
	\newcommand{\ee}{\end{equation}}
\newcommand{\ba}{\begin{eqnarray}}
	\newcommand{\ea}{\end{eqnarray}}
\newcommand{\bref}[1]{(\ref{#1})}

\newcommand{\bi}[1]{\bibitem{#1}}\newcommand{\lab}[1]{\label{#1}}

\newcommand{\bsub}{\begin{subequations}}                      
	\newcommand{\esub}{\end{subequations}}

\begin{document}

	\preprint{APS/123-QED}
	
	\title{
		 Bright-soliton frequency combs and dressed states in  $\chi^{(2)}$ microresonators
	}
	\author{D.N. Puzyrev}
	\author{V.V. Pankratov} 
	\author{A. Villois} 
	\author{D.V. Skryabin}   
	\email{d.v.skryabin@bath.ac.uk} 
	\affiliation{Department of Physics, University of Bath, Bath, BA2 7AY, UK}
	\date{today}
	
	\begin{abstract}
We present a theory  of the frequency comb generation in the high-Q ring microresonators with quadratic nonlinearity and normal dispersion and demonstrate that the naturally  large  difference of the repetition rates 
at the fundamental and 2nd harmonic frequencies supports 
a family of the bright soliton frequency combs providing the parametric gain is moderated by tuning  
the index-matching parameter to exceed the repetition rate difference by a significant factor. 
This factor equals the sideband number associated with the high-order phase-matched sum-frequency process.
The theoretical framework, i.e., the dressed-resonator method, to study the frequency conversion and comb generation is formulated by including the sum-frequency nonlinearity into the definition of the resonator spectrum. 
The Rabi splitting of the dressed frequencies leads to the four distinct 
parametric down-conversion  conditions (signal-idler-pump photon energy conservation laws).
The parametric instability tongues associated  with the generation of the sparse, i.e., Turing-pattern-like, frequency combs with varying repetition rates are analysed in details. The sum-frequency matched sideband 
exhibits the optical Pockels nonlinearity and strongly modified dispersion, 
which limit the soliton bandwidth and also play a distinct role in the Turing comb generation. Our methodology and data highlight the analogy between the driven multimode resonators and the photon-atom interaction.
	\end{abstract}
	\maketitle

	\tableofcontents


\section{Introduction}\lab{intro}
Ring microresonators break through the traditional barriers of frequency conversion in 
terms of the power efficiency, generated bandwidth, and compactness~\cite{al,rev3}.
Together with the rise of the microresonator frequency conversion, 
the Kerr-soliton frequency combs are reaching unprecedented for optical solitons 
levels of practical relevance~\cite{al,rev3}. 

Second-order, $\chi^{(2)}$, nonlinearity is a viable alternative to the Kerr one. 
It provides a much stronger nonlinear response but comes with the caveats of the need 
to care about the phase and group velocity matching to take full advantage of it.  
Refs. \cite{prl0,prl1,prl2,prl3} have been among the first ones 
to  demonstrate frequency conversion in the  high-quality factor 
whispering gallery microresonators with quadratic nonlinearity. 
Since then, this area has made a significant progress, see Ref.~\cite{ingo} for the few-year old overview, and Refs.~\cite{mian,micro5,japan,fins,jan1,jan2,miro,bru} for some of the recent experimental contributions.  It is also important to mention the work on the  $\chi^{(3)}$ dominant frequency conversion in the significantly mismatched $\chi^{(2)}$ resonators, see, e.g.,~\cite{micro9,micro4,micro3,gaeta}, and  Appendix~\ref{ap2} for the relative weighting of the $\chi^{(2)}$ and $\chi^{(3)}$ effects.
Material wise, lithium-niobate (LN) remains the best-established platform choice for the small-footprint $\chi^{(2)}$ photonics~\cite{mian}, with the  silicon~\cite{sn} and aluminium~\cite{bru} nitrides gaining grounds with  accelerating pace.

The whispering gallery system considered below achieves  finesses $\cF\sim 10^4$.
The high finesse, is an important prerequisite  for achieving
conditions when the rate of the sideband generating photon energy exchange 
driven by the sum-frequency nonlinear terms exceeds the loss rate~\cite{tang3,tang2} and 
the parametric gain rate~\cite{prr}, which corresponds to the strong coupling regime between the
$\omega$ and $2\omega$ photons. 
As we have reported recently~\cite{prr} and investigate in depth below, this makes the frequency conversion and soliton generation mechanisms to 
depart significantly from what has been known about these effects in the relatively
low finesse resonators, $\cF\sim 10^2$, which often have
no resonator-feedback at one of the generated harmonics, 
see, e.g., Refs.~\cite{mash,mar} for an overview. A couple of most obvious and important features
of the high-$\cF$ resonators is the channelling of the parametric gain into the tongue-like instability domains~\cite{prr} and the strong-coupling associated with the dressed states~\cite{tang3}
and polaritons~\cite{prr}.

Experimental results on the $\chi^{(2)}$-driven microresonator solitons are limited for now
by the outstanding recent report by Bruch 
and colleagues \cite{bru} on the solitons due to parametric down-conversion (PDC) in the aluminium-nitride microring with the finesse $\simeq 10^3$.  The numerical data reported
in Ref.~\cite{bru} show the exponentially localised pulse in the half-harmonic field and the de-localised wave-form in the pump. 
Ref.~\cite{bru} poses several problems, in particular, what are the physical mechanisms facilitating the compensation of the large group velocity, i.e., the repetition rate, difference between the pump and half-harmonic fields allowing for the soliton to form, and if having a better shaped pulse in the pump field is possible. Identifying conditions for the multi-colour multi-pulse modelocked solitons to compensate the large group velocity differences is one of the classic problems~\cite{dunn,preold}, that needs to be addressed in the context of the high-Q microresonators (see Ref.~\cite{icfo} for the no-resonator, i.e., bulk propagation, analysis). 

Results on the bright PDC solitons in the resonators  with the group velocity offset published two decades ago~\cite{preold} 
provided a conceptual answer, that the compensation of the group velocity difference is achieved via 
the balancing  interplay between  the dissipative and nonlinear effects. However,  Ref.~\cite{preold}  was published well before the frequency combs and solitons in  the
micro-resonators have become possible. Therefore, it could not anticipate a combination of the small sizes 
and high Q-factors of the modern day devices, which leads to the strong quantisation of the spectra of the operators underpinning the frequency conversion and soliton formation processes. 
The model that fully reflects on these aspects has been recently presented in Ref.~\cite{josab}, which introduction also includes sufficiently comprehensive coverage of the recent and historic work  on the dissipative $\chi^{(2)}$ solitons in resonators.

Following our recent work on the theory of $\chi^{(2)}$ microresonators~\cite{josab,av,ol,jan1,prr}, we present here the latest findings obtained in the second harmonic generation (SHG) setting. One of the  prime results included below is the demonstration of the bright soliton pulses in a  microresonator which has the large repetition rate difference between the fundamental, $\om_p$, and 2nd harmonic frequencies, $2\om_p$, see Figs.~\ref{sol2}-\ref{sol3}. 

The resonator considered here has the $\simeq 20$GHz repetition rate and the $1$GHz rate difference, which  implies that the linear $\om_p$ and $2\om_p$ pulses would be on the opposite sides of the resonator only after about seventy round-trips. The dispersion of the resonator is normal and its deviation from the bulk dispersion is insignificant. The dispersion value is $\simeq -100$kHz so that the linear pulse becomes
twice as broad after $10^5$ round-trips, and hence the repetition-rate difference is 
by far the most dangerous for the bright soliton modelocking, which nevertheless will be shown to exist across the broad and practical range of the system parameters.

The soliton combs reported below represent a pair of the $\om_p$  and $2\om_p$ modelocked, and hence the repetition rate locked, bright pulses which existence and properties are derived and interpreted by 
examining the details of how the comb teeth, i.e., the frequency sidebands, are generated and interact via the sum-frequency and PDC nonlinear mixing processes. The detailed understanding of this problem has become possible thanks to the dressed resonator theory~\cite{prr}. 
The content of what follows is much wider than just reporting the soliton modelocking, and
 it is now useful to give it a brief overview. 

\section{Content and results overview}\lab{sec0}
The coupling between the sidebands, i.e., the resonator mode pairs or the comb teeth, 
underpins the formation of any frequency combs and modelocking in optical resonators. 
Below we consider the microresonator  2nd harmonic generation, where the  sideband coupling mechanisms embrace the complex interplay of the parametric  process that couples the $\mu$ and $-\mu$  sidebands with the two sum-frequency processes.  One is responsible for the coupling between  the $\mu$ sidebands of the fundamental and 2nd harmonic, while the other does the same but for the $-\mu$ sidebands. The sum-frequency processes can be phase-matched for the select sidebands with the positive and negative numbers,  $\mu=\mu_*$ or $\mu=-\mu_*$, where $\mu_*$ is the ratio between the phase-mismatch parameter
and the repetition rate difference. 
The key resonator parameters and characteristics, including $\mu_*$, 
are illustrated in Fig.~\ref{fspectr}, and defined in Tables~\ref{t1} and \ref{t2}.

If the microresonator is tuned to operate away from the $\mu=0$ phase-matching, 
then the sum-frequency sideband coupling rate exceeds the parametric gain.  It creates an opportunity to redefine the resonator spectrum by including the sum-frequency associated nonlinear terms. The new spectrum, i.e., the dressed spectrum, can be calculated analytically, bringing  a close analogy with the dressing of the atomic transitions  and the Rabi theory~\cite{tang3,prr,boyd}. 

The condition of the maximal parametric gain~\cite{byer0,byer1,st0}, 
\be
2\hbar\om_p=\hbar\omega_{\text{signal}}+\hbar\omega_{\text{idler}},
\ee 
should then be redefined
using the dressed frequencies~\cite{prr}. The Rabi splitting 
leads to the four distinct PDC conditions, see Eq.~\bref{z0}. 
Knowing them allows understanding the complex structures  
of the parametric instability tongues and the Turing-pattern like frequency combs across the parameter space spanned by the pump laser frequency, $\om_p$, 
and the intra-resonator power.

The dressed modes around either $\mu_*$ or $-\mu_*$ exhibit dispersion with  the inverted signs and the  values significantly exceeding the bare-resonator ones. Therefore, only by making $\mu_*$ to be sufficiently large, i.e., by taking it away from the soliton spectral core, creates enough of the modal bandwidth around $\mu=0$ with the  low  dispersion allowing the formation of the two-color bright soliton pulses. 
The large values of  $\mu_*$ are achieved by making the frequency mismatch parameter between the fundamental and 2nd harmonic modes to exceed the repetition rate difference by a significant, $\gg 1$, factor, which works out to be the $\mu_*$ itself.

\section{Linearised sideband equations and cw-state}\lab{sec2}
We assume the  intra-resonator electric fields of the fundamental and 2nd 
harmonic to be
\be
\psi_\f e^{iM\vta-i\om_p t}+c.c.,~
\psi_\s e^{i2M\vta-i2\om_p t}+c.c.,
\lab{field}
\ee 
where $|\psi_{\f,\s}|^2$  have units of power~\cite{josab}, see Appendix~\ref{ap1}.  
$M$ and $2M$ are the absolute (physical) mode numbers with  
frequencies $\om_{0 \f}$ and  $\om_{0 \s}$. $\vta\in (-\pi,\pi]$ is the angular coordinate and $t$ is time.
$\om_p$ is the pump laser frequency tunable around $\om_{0 \f}$, so that,
\be
\delta=\om_{0\f}-\om_p
\lab{de}
\ee
is the respective pump detuning.

The envelopes of the fundamental, $\psi_\f$, and second harmonic, $\psi_\s$, 
are expressed via their mode expansions as 
\be
\psi_\zeta=\psi_{0\zeta}(t)+\sum_{\mu>0}\left(\psi_{\mu\zeta}(t)e^{i\mu\vta}
+\psi_{-\mu\zeta}(t)e^{-i\mu\vta}\right),
\lab{b0}
\ee 
with $\zeta=\f,\s$, and $\mu$ is an integer characterising the relative mode number. The resonator frequencies associated with $\psi_{\pm\mu\zeta}$  are
\be
\om_{\pm\mu\zeta}=\om_{0\zeta}\pm\mu D_{1\zeta}+\tfrac{1}{2}\mu^2 D_{2\zeta},
\lab{om0}
\ee
 where $D_{1\zeta}/2\pi$ are the repetition rates (free spectral ranges) and 
$D_{2\zeta}$ are the dispersions, see Fig.~\ref{fspectr} for a schematic illustration. 
The  photon angular momenta  corresponding to $\om_{\pm\mu\f}$ and $\om_{\pm\mu\s}$ are
\be
\hbar(M\pm\mu),~\text{and}~ \hbar(2M\pm\mu).\nn
\ee 

\begin{figure}[t]
	\centering
	\includegraphics[width=0.49\textwidth]{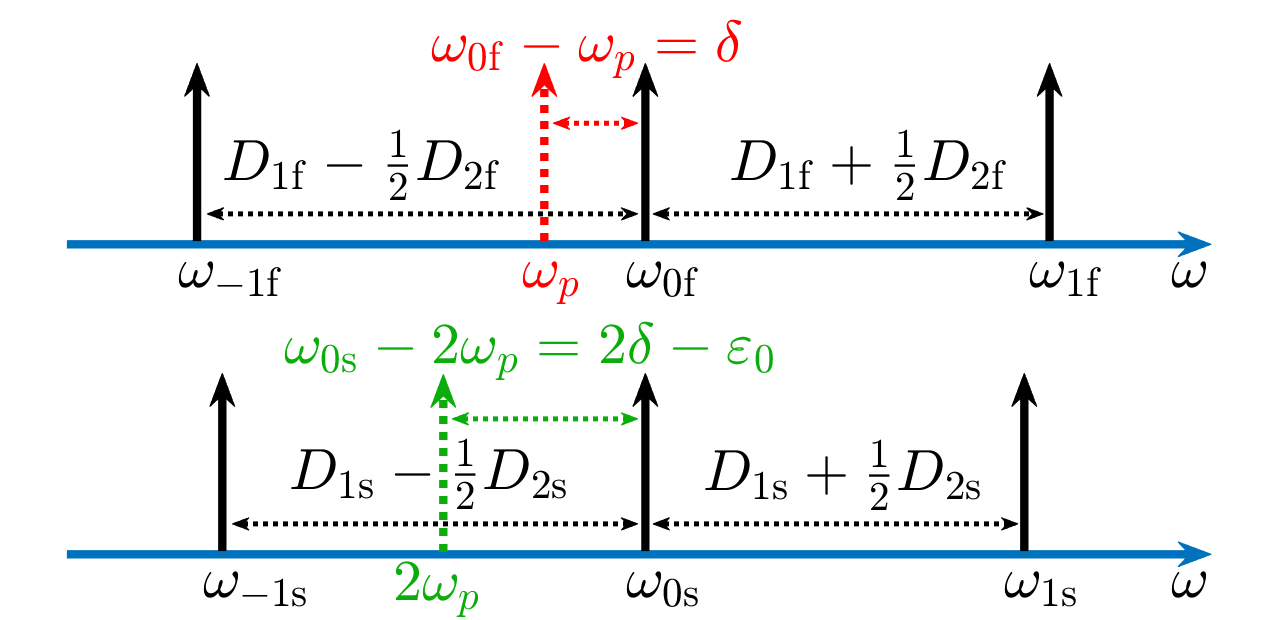}
	\caption{A schematic illustration of the spectrum of the linear, i.e., bare resonator, $\Omega=0$, around the pump laser frequency, $\om_p$, and its 2nd harmonic, $2\om_p$.  The frequency mismatch parameter $\ep_0$ is tunable and could be comparable with $D_{1\zeta}$.}
	\lab{fspectr}
\end{figure}

\begin{table}[t]
	\begin{ruledtabular}
		\begin{tabular}{l}
			Linewidths: $\kappa_\f/2\pi=1$MHz, $\kappa_\s/2\pi=4$MHz 
			\\ \hline \\
			Repetition rates: $D_{1\f}/2\pi=21$GHz, $D_{1\s}/2\pi=20$GHz
			\\ \hline \\
			Dispersions: $D_{2\f}/2\pi=-100$kHz, $D_{2\s}/2\pi=-200$kHz
			\\ \hline \\
			Nonlinear coefficients: $\gamma_{\f,\s}/2\pi=300\text{MHz}/\sqrt{\text{W}}$
		\end{tabular}
	\end{ruledtabular}
	\caption{The resonator  parameters characteristic for  a bulk-cut  LiNbO$_3$ resonator  pumped at $1065$nm \cite{jan1,josab,prr}. $D_{2\f,2\s}<0$ correspond to the normal dispersion.
	}
	\lab{t1}
\end{table}

Physical values of the parameters implemented throughout this work are shown in Table~\ref{t1}, and
are typical for a bulk-cut LiNbO$_3$ resonator. 
Detunings of the resonator frequencies $\om_{\pm\mu\zeta}$ from the pump laser frequency, $\om_p$,
and its 2nd harmonic, $2\om_p$, are  $\om_{\pm\mu \f}-\om_p$
and $\om_{\pm\mu \s}-2\om_p$. A Galilean transformation to the reference frame rotating with the 
rate $D_{1}/2\pi$, 
\be
\ta=\vta-D_{1}t,
\lab{gal}
\ee 
converts these detunings to 
\be
\begin{split}
	\vD_{\pm\mu\f}& =(\om_{\pm\mu\f}-\om_p)\mp\mu D_{1}\\ 
	&=\delta
	\pm\mu(D_{1\f}-D_{1})+\tfrac{1}{2}D_{2\f}\mu^2,\\ 
	\vD_{\pm\mu\s}&=(\om_{\pm\mu\s}-2\om_p)\mp\mu D_{1}\\&=2\delta-\ep_0\pm\mu(D_{1\s}-D_{1})+\tfrac{1}{2}D_{2\s}\mu^2,
\end{split}
\lab{detu}
\ee
where $\ep_0=2\om_{0\f}-\om_{0\s}$ is the $\mu=0$ frequency mismatch parameter.
Setting
\be
D_1=D_{1\f},
\ee
we get an immediate access to the difference of the repetition rates, $D_{1\s}-D_{1\f}$,
inside $\vD_{\pm\mu\s}$.

\begin{figure}[t]
	\centering
	\includegraphics[width=0.35\textwidth]{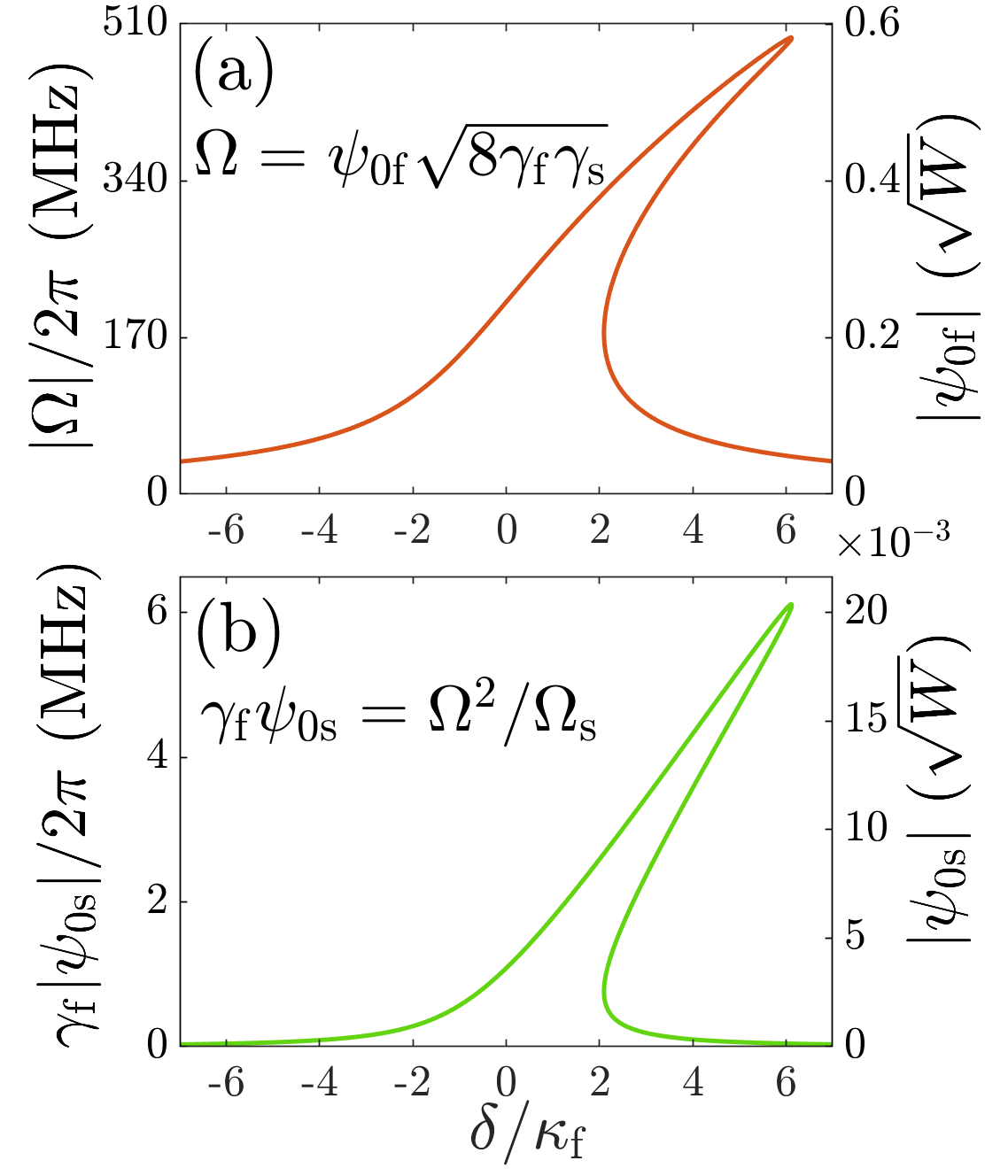}
	\caption{CW state of the fundamental~(a) and 2nd harmonic~(b), see Appendix~\ref{ap3}. Pump laser power is $\cW= 0.1$mW, see Eq.~\bref{cw8}. Mismatch parameter 
		$\varepsilon_0 / 2 \pi = -5$GHz.
	}
	\lab{fcw}
\end{figure}

If $c$ is the vacuum speed of light, $n_{M+\mu}$ is the effective refractive index of the mode $M+\mu$, and $R$ is the resonator radius, then 
\be
\om_{\mu\f}=\frac{c}{R}\frac{(M+\mu)}{n_{M+\mu}},~\om_{\mu\s}=\frac{c}{R}\frac{(2M+\mu)}{n_{2M+\mu}}.
\lab{refin}
\ee 
Hence,
\be
\ep_0=2\om_{0\f}-\om_{0\s}=
2M \frac{c}{R}
\left[\frac{1}{n_{M}}-\frac{1}{n_{2M}}\right],
\lab{en22}
\ee
and requiring $\ep_0=0$, 
yields the anticipated  index matching condition, $n_{M}=n_{2M}$. 
Refractive index, and hence $\ep_0$, can be fine tuned by, e.g., temperature 
or electro-optic controls. 

CW state of the resonator operation implies   
$\psi_{\zeta}=\psi_{0\zeta}=\text{const}_\zeta$, i.e., $\p_t \psi_{0\zeta}=0$, see Fig.~\ref{fcw} and Appendix~\ref{ap3}.
During the initial stage of the comb development, the  sidebands evolve and grow on top of the  undepleted cw-state, so that the field envelopes can be sought as the cw plus small perturbations, with the latter taken as a superposition of the   resonator modes,
\be
\psi_\zeta(t,\ta)=\psi_{0\zeta}+\sum_{\mu\ge 0}\left[\wt\psi_{\mu\zeta}(t)e^{i\mu\ta}
+\wt\psi_{-\mu\zeta}^*(t)e^{-i\mu\ta}\right].
\lab{anz}
\ee
Here $\wt\psi_{\pm\mu\zeta}$ are the amplitudes of the growing sidebands. Complex conjugation of $\wt\psi_{-\mu\zeta}$ was introduced to make the equations to follow less cluttered. 
If Eq.~\bref{b0} could be referred to as the mode expansion in the bare resonator 
representation, then Eq.~\bref{anz} is a step towards the dressed resonator theory.

Substituting Eq.~\bref{anz} to the envelope equations~\bref{mm}, and linearising for small sidebands,
we find that $\wt\psi_{\mu\zeta}(t)$ are driven by the sideband combinations with the matched net momenta, see Appendix~\ref{ap4}. The resulting equations  are
\be
\begin{split}
	&i\p_t \wt\psi_{\mu\f} =(\vD_{\mu\f}-i\tfrac{1}{2}\kappa_\f)\wt\psi_{\mu\f}-\gamma_{\f}
	(\underline{\wt\psi_{\mu\s}\psi_{0\f}^*}+
	\underline{\underline{\wt\psi_{-\mu\f}\psi_{0\s}}}),\\ 
	&i\p_t \wt\psi_{\mu\s}=(\vD_{\mu\s}-i\tfrac{1}{2}\kappa_\s)\wt\psi_{\mu\s}
	-2\gamma_{\s}\underline{\wt\psi_{\mu\f}\psi_{0\f}},\\
	&i\p_t \wt\psi_{-\mu\f} =(-\vD_{-\mu\f}-i\tfrac{1}{2}\kappa_\f)\wt\psi_{-\mu\f}
	+\gamma_{\f}(\underline{\wt\psi_{-\mu\s} \psi_{0\f}}+
	\underline{\underline{\wt\psi_{\mu\f} \psi_{0\s}^*}}),\\ 
	&i\p_t \wt\psi_{-\mu\s}=(-\vD_{-\mu\s}-i\tfrac{1}{2}\kappa_\s)\wt\psi_{-\mu\s}+2\gamma_{\s} 
	\underline{\wt\psi_{-\mu\f}\psi_{0\f}^*}.  
\end{split} 
\lab{kk}
\ee
Here,  
$\kappa_\zeta$ are the loaded linewidth parameters, and 
$\gamma_\zeta$ are the nonlinear coefficients measured in Hz/W$^{1/2}$~\cite{josab}, see Table~\ref{t1}. 

Eq.~\bref{kk} are linear in the approximation of the undepleted cw-state, and they 
represent one of the fundamental models in nonlinear optics expressing the interplay of  the PDC 
and sum-frequency processes. The sum-frequency terms are underlined once and 
the parametric ones, describing conversion to the $\pm\mu$ sidebands (photon-pair generation),
are underlined twice. 
The momentum conservation laws corresponding to the parametric conversion and 
the two sum-frequency processes  are 
\be
\hbar (M+\mu)+\hbar(M-\mu)=\hbar 2M,
\lab{mom1}
\ee 
and 
\be
\hbar(M\pm\mu)+\hbar M=\hbar(2M\pm\mu),
\lab{mom2}
\ee 
respectively. 
 
The eigenvalues and eigenvectors of the matrix acting on the vector $(\wt\psi_{\mu\f},\wt\psi_{\mu\s},\wt\psi_{-\mu\f},\wt\psi_{-\mu\s})^T$ in the right-hand side  of Eq.~\bref{kk} are known in the explicit form  in two cases: (i)~for
$\vD_{\mu\f}=\vD_{-\mu\f}$,  $\vD_{\mu\s}=\vD_{-\mu\s}$ and $\kappa_\f=\kappa_\s$, see, e.g., Refs.~\cite{lug,bur,wabol}, and (ii)~for $\kappa_\f\ne\kappa_\s$ and $\vD_{\pm\mu\zeta}=0$, see
Ref.~\cite{drum}. While $\vD_{\mu\f}=\vD_{-\mu\f}$ is satisfied exactly, and $\kappa_\f=\kappa_\s$ could be assumed,  the $\vD_{\mu\s}=\vD_{-\mu\s}$ condition is typically far from being true. In most of the practical cases, the repetition rate difference, $\vD_{\mu\s}-\vD_{-\mu\s}=2\mu(D_{1\s}-D_{1\f})$, creates one of the dominant frequency 
scales in Eqs.~\bref{detu}, which can be comparable only to $\ep_0$. 

\begin{figure}[t]
	\centering
	\includegraphics[width=0.48\textwidth]{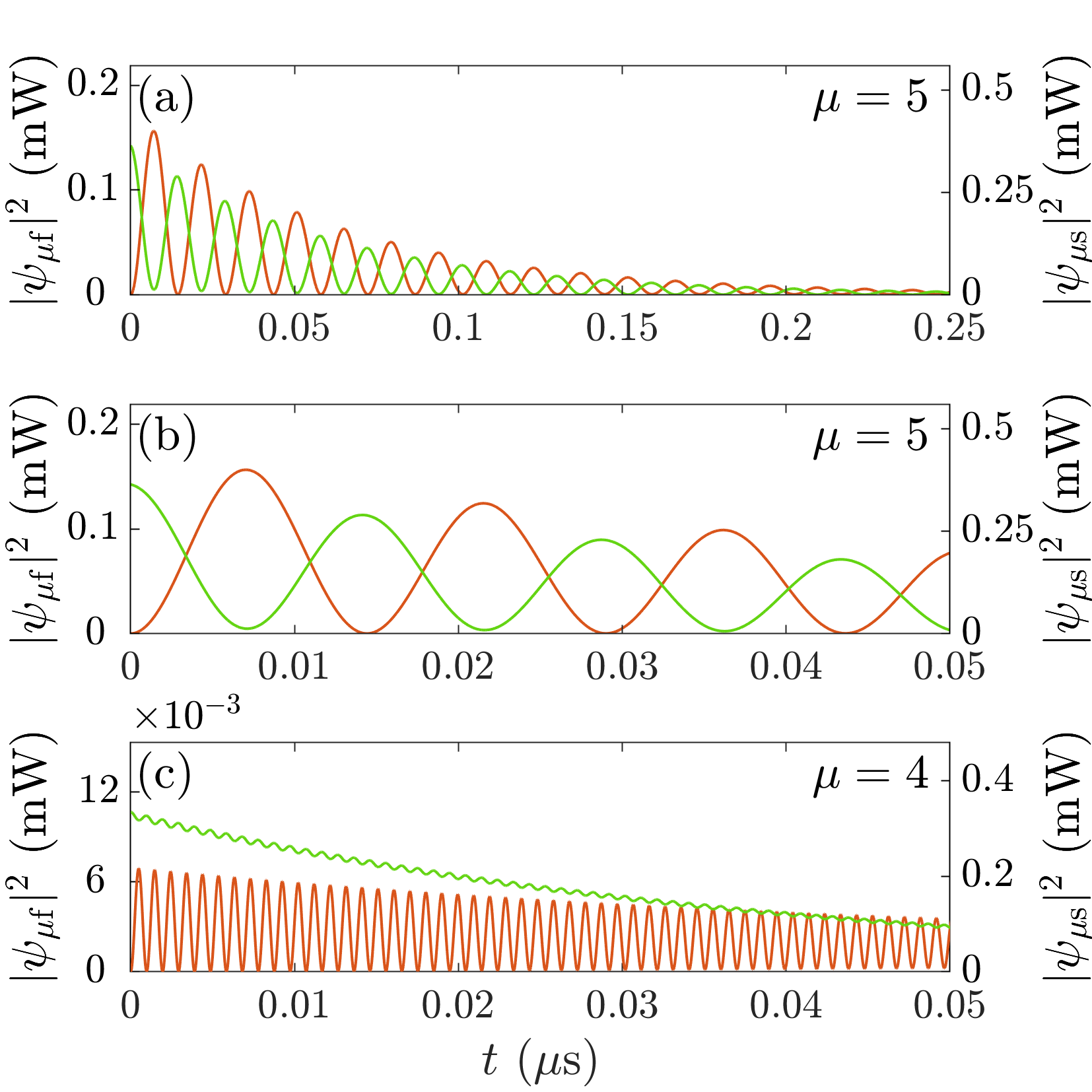}
	\caption{
		(a)~The sum-frequency matched Rabi oscillations near to the $\mu=5$  avoided crossing  (red - fundamental, left axis; green - second harmonic, right axis), $\Omega_5\approx|\Omega|\approx 2\pi\times 68$MHz. (b)~is as (a)~but it is plotted  over the time interval as is used in~(c). (c)~The mismatched Rabi oscillations for the $\mu=4$ sidebands, $\Omega_4\approx |\vD_4|\approx 2\pi\times 1$GHz. Parameters: (a,b)~$\delta=15.15\kappa_\f$, $\ep_0/2\pi = -5$GHz, $\cW = 1.7$mW (pump power), $|\psi_{5\s}|^2=0.36$mW (sideband power) at $t=0$. (c)~$\delta=6.45\kappa_\f$, $\ep_0/2\pi= -5$GHz, $\cW = 2.11$mW, $|\psi_{4\s}|^2=0.33$mW at $t=0$. 
	}
	\lab{frabi}
\end{figure}

\section{Rabi frequency, parametric gain and linewidth}\lab{sec3}
To address  the problem at hands, it is important to understand the balances between the characteristic frequency scales  implicated in Eq.~\bref{kk}. 
$\psi_{0\s}$, i.e., the cw 2nd harmonic amplitude, can be controlled  
by tuning the frequency mismatch, $\ep_0$, so that the parametric gain coefficient, $\gamma_\f\psi_{0\s}$, can be made to be much less than the sum-frequency associated rate of the energy exchange, $\gamma_\zeta\psi_{0\f}$, see Fig.~\ref{fcw} and compare the frequency scales along the vertical axes.

If the aim is to create the modal bandwidth sufficient for the soliton generation at both harmonics, then the moderation of the parametric gain is useful since it would then keep the energy of the 2nd harmonic pulse
under control, and, hence, could be expected to ease, for the stronger fundamental pulse, the task of synchronising its repetition rate with the 2nd harmonic.  

The sum-frequency driven energy exchange is illustrated in Fig.~\ref{frabi}, where one can see the 
fast anti-phase oscillations of the $\om_{\mu\f}$ and $\om_{\mu\s}$ 
sidebands. The frequency of the oscillations is 
much larger than the decay rates, so that, in the leading order, the first two equations in  Eqs.~\bref{kk} can be approximated with~\cite{prr,pral}
\be
i\p_t\begin{bmatrix} \wt\psi_{\mu\f}\\ \wt\psi_{\mu\s}\end{bmatrix}
\approx \begin{bmatrix} 
	\vD_{\mu\f}&-\gamma_{\f}
	\psi_{0\f}^*\\
	-2\gamma_{\s}\psi_{0\f}	&\vD_{\mu\s}
\end{bmatrix}\begin{bmatrix} \wt\psi_{\mu\f}\\ \wt\psi_{\mu\s}\end{bmatrix}+\dots.
\lab{re1}
\ee
The second pair of equations resemble the above  but with $\mu\to-\mu$.
Parametric gain and losses should then come 
in the next to leading order, suggesting  the development of a perturbation theory.
The Rabi theory, well known in the semi-classical atom-photon
interaction~\cite{boyd}, is an obvious and best suited methodology 
to describe solutions of Eq.~\bref{re1}. 
The Rabi  formalism was also previously applied for the resonator-free 
sum-frequency generation model~\cite{adi}.
\begin{table}[t]
	\begin{ruledtabular}
		\begin{tabular}{ll}
			Rabi frequency& $\Omega=\psi_{0\f}\sqrt{8\gamma_\f\gamma_\s}$
			\\ 
			&~~~$\approx\psi_{0\f}\times 0.8~$GHz/$\sqrt{\text{W}}$
			\\ \hline \\
			Rabi detuning& $\vD_\mu=\om_p+\om_{\mu\f}-\om_{\mu\s}$
			\\ \hline \\
			Sum-frequency mismatch &$\ep_\mu=\om_{0\f}+\om_{\mu\f}-\om_{\mu\s}$
			\\ \hline \\
			Effective Rabi frequency & $\Omega_\mu=\sqrt{\vD_\mu^2+|\Omega|^2}$
			\\ \hline \\
			PDC frequency mismatch &$\ep_\mu^{(j_1j_2)}=\wt\om_{\mu\f}^{(j_1)}+\wt\om_{-\mu\f}^{(j_2)}-2\om_p$	
			\\ \hline \\
			Strong-coupling condition  & $\kappa_\zeta\ll |\Omega|\ll 8|\ep_0|$
			\\ \hline \\
			Sum-frequency matching & $\ep_{\pm\mu_*}=0,$ $\mu_*\approx|\ep_0|/|D_{1\f}-D_{1\s}|$
		\end{tabular}
	\end{ruledtabular}
	\caption{Definitions of the key parameters, which complement the ones illustrated in Fig.~\ref{fspectr} and listed in Table~\ref{t1}. $\om_p$~is the laser photon frequency. 
		$\om_{\mu\zeta}$~and  $\wt\om_{\mu\zeta}^{(j)}$ are the bare and dressed resonator frequencies, respectively ($j_{1,2}=1,2,3,4$; $\zeta=\f,\s$).}
	\lab{t2}
\end{table}

The frequency of the oscillations in Fig.~\ref{frabi} would then be the effective Rabi frequency,
\be
\Omega_\mu=\sqrt{\vD_\mu^2+|\Omega|^2}.
\lab{ra00}
\ee
It is controlled by the difference of the sideband detunings, i.e.,
the Rabi detuning, 
\be
\vD_\mu=\vD_{\mu\f}-\vD_{\mu\s},
\lab{del}
\ee 
and by the coupling coefficient, i.e., by the off-diagonal terms in Eq.~\bref{re1}, 
characterised by the complex Rabi frequency, $\Omega$,
 \be
\Omega=\psi_{0\f}\sqrt{8\gamma_{\f}\gamma_{\s}}.
\lab{ws10}
\ee 

The complex 2nd harmonic amplitude can also be expressed via $\Omega$ and the auxiliary complex frequency $\Omega_\s$,
 \be
 \gamma_{\f}\psi_{0\s}=\frac{\Omega^2}{
		\Omega_{\s}},~
\Omega_\s=8(2\delta-i\tfrac{1}{2}\kappa_\s)-8\ep_0,
\lab{os}
\ee  
see Appendix~\ref{ap3}.

$|\Omega|/2\pi\sim 10^2$MHz gives $|\psi_{0\f}|^2\lesssim 1$W, which would be typical 
inside the resonator. Then, for 
$\kappa_\zeta/ 2\pi\sim 1$MHz  we have $|\Omega|\gg\kappa_\zeta$, i.e., the Rabi flops are indeed much faster than the decay rate. 
The frequency scale associated with the parametric gain is set by 
\be
\gamma_\f|\psi_{0\s}|=\frac{|\Omega|^2}{|\Omega_\s|}.
\lab{pa1}
\ee 
Arranging $|\ep_0|$ to be  close or larger than the repetition rate difference,
\be
\frac{|\ep_0|}{2\pi}\gtrsim \frac{|D_{1\f}-D_{1\s}|}{2\pi}\simeq 1\text{GHz},
\lab{in1}
\ee
makes 
\be
\frac{1}{|\Omega|}\frac{|\Omega|^2}{|\Omega_\s|}\approx\frac{|\Omega|}{8|\ep_0|}\ll 1.
\lab{pa2}
\ee
 Hence,  the Rabi frequency is also much larger than the parametric gain rate. 
Thus, both the linewidth  and the parametric terms are small relative to the right-hand side of Eq.~\bref{re1}, and the strong-coupling (SC) condition~\cite{prr}, 
\be
\kappa_\zeta\ll|\Omega|\ll|\Omega_\s|,
\lab{sc0}
\ee 
is satisfied. Ref.~\cite{tang3}
reported measurements of the $\chi^{(2)}$ Rabi splitting ($\sim 1$GHz for the laser power $\cW=80$mW) in the AlN resonators with $|\Omega|/\kappa_\zeta\sim 1$, which should be well out-performed 
by the bulk resonators considered here. The~$\mu=0$ Rabi oscillations in the $\chi^{(2)}$ resonators were looked at in Ref.~\cite{car},  well before the multimode high-Q $\chi^{(2)}$ microresonators  have been demonstrated.

The notations related to the Rabi theory and also the key quantities used below to characterise the matching conditions for the sum-frequency and parametric processes are summarised in Table~\ref{t2}.

\begin{figure}[t]
	\centering
	\includegraphics[width=0.49\textwidth]{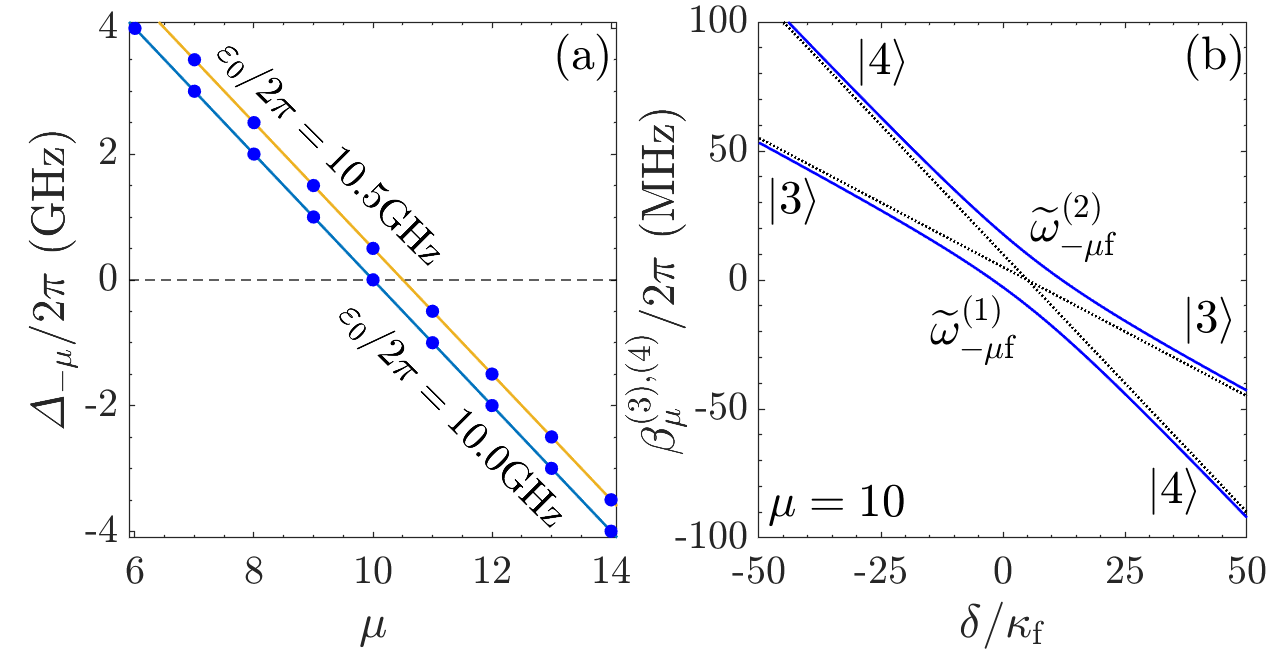}
	\caption{(a)~Frequency matching for the sum-frequency generation: The Rabi detuning $\vD_{-\mu}=\om_p+\om_{-\mu\f}-\om_{-\mu s}$ vs the sideband order number $\mu$, $\delta=5\kappa_\f$. 
  $\ep_0/2\pi=10.5$GHz 	and $\ep_0/2\pi=10$GHz correspond to the mismatched and  near-matched ($\mu=\wh\mu=10$) cases, respectively. 	(b)~Avoided-crossing diagram: The dressed frequencies   vs $\delta$ for $|\Omega|=20\kappa_\f$. The straight lines correspond to $\Omega=0$. }
	\lab{fcross}
\end{figure}

\section{Sum-frequency matching}\lab{sec4}

While the cascade of the sum- and difference-frequency events engaging a sequence of $\mu$'s is critical for the generation of the fully blown combs, see Section~\ref{sec9}, the frequency matching for the one-step sum-frequency process in Eq.~\bref{mom1} is also very important and should be analysed in more details. As it is well known for the coupled oscillators, the full periodic power transfer between $\wt\psi_{\mu\f}$ and $\wt\psi_{\mu\s}$ is ensured by minimizing the Rabi detuning, i.e., $\vD_\mu\to 0$, $\Omega_\mu\to|\Omega|$, see Eq.~\bref{ra00}. 

Here, we are dealing with a system possessing two different effective Rabi frequencies $\Omega_\mu$ and $\Omega_{-\mu}$, and the respective Rabi detunings, $\vD_{\pm\mu}$,  can be re-expressed as
\be
	\vD_{\pm\mu}=\om_p+\om_{\pm\mu\f}-\om_{\pm\mu\s}.
	\lab{vd}
\ee
If one of $\vD_{\pm\mu}=0$ is resolved by an integer, i.e., 
\be
\vD_{\wh\mu}=0,~\text{or}~\vD_{-\wh\mu}=0,~\wh\mu\in\mathbb{Z},~\wh\mu>0,
\lab{ha}
\ee 
it implies the exact matching for one of the two co-existing 
sum-frequency processes. The examples of the exact matching between
$\om_{-\mu\f}$ and $\om_{-\mu\s}$ for $\mu=10$ and of the mismatched case are shown in Fig.~\ref{fcross}(a). 
In the mismatched case, the  SC condition, Eq.~\bref{sc0}, is perfectly satisfied, but the power transfer during the Rabi flops is significantly reduced, cf., Figs.~\ref{frabi}(b) and (c). 

\begin{figure}[t]
	\centering
	\includegraphics[width=0.49\textwidth]{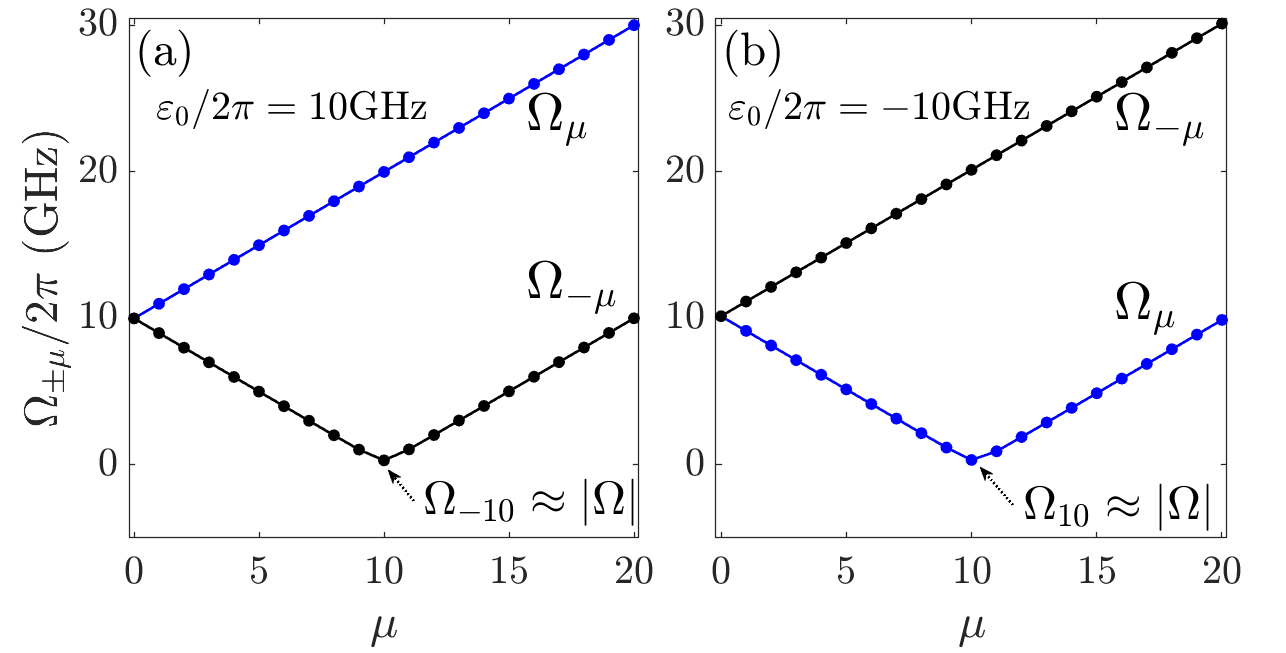}
	\caption{$\Omega_\mu$ and $\Omega_{-\mu}$ Rabi frequencies vs $\mu$. (a) $\ep_0/2\pi=10.5$GHz, and (b) $\ep_0/2\pi=-10.5$GHz, $\delta=14\kappa_\f$, $|\Omega|=300\kappa_\f$. $\vD_{-\mu}=\om_p+\om_{-\mu\f}-\om_{-\mu\s}$ is nearly matched for $\mu= 10$ in (a), and $\vD_{\mu}=\om_p+\om_{\mu\f}-\om_{\mu\s}$ is in (b). 
	}
	\lab{frabi0}
\end{figure}

If the real positive $\mu_*$ solves one of the $\vD_{\pm\mu}=0$ equations, i.e., 
\be
\vD_{\mu_*}=0,~\text{or}~\vD_{-\mu_*}=0,~\mu_*\in\mathbb{R},~\mu_*>0,
\ee
then an  integer, or two,  nearest to $\mu_*$ 
provide the sidebands order best complying with the frequency matching. 
To find $\mu_*$, we  introduce a new parameter 
\be
\ep_{\pm\mu}=\om_{0\f}+\om_{\pm\mu\f}-\om_{\pm\mu\s},
\lab{x1}
\ee 
cf., Eq.~\bref{vd}, so that,
\be
\vD_{\pm\mu}=\ep_{\pm\mu}-\delta.
\lab{x2}
\ee
We note that $\ep_{\pm\mu}$
depend only on the resonator geometry and refractive index, 
cf.,~Eq.~\bref{en22}. 
Using Eq.~\bref{om0}, we express $\ep_{\pm\mu}$  as
\be
\ep_{\pm\mu}=\ep_0\pm\mu(D_{1\f}-D_{1 \s})+\tfrac{1}{2}\mu^2 (D_{2\f}-D_{2 \s}),
\ee
Relative smallness of the dispersion, i.e., of the $\mu^2 D_{2\zeta}$, and $\delta$ terms,
provides an excellent approximation for $\mu_*$,
\be
\mu_*\approx\frac{|\ep_0|}{|D_{1\f}-D_{1 \s}|}.
\lab{epe}
\ee

If  $\ep_0>0$, then $\mu_*$ solves $\vD_{-\mu}=0$, see Fig.~\ref{fcross}(a), and if 
$\ep_0<0$, then $\mu_*$ solves $\vD_{\mu}=0$. This point is further facilitated
in Fig.~\ref{frabi0}, where we plot $\Omega_\mu$ and $\Omega_{-\mu}$ vs $\mu$
for different signs of $\ep_0$. For $\ep_0>0$, $\Omega_{-\mu}$ has minimum at $\mu=\mu_*\approx\wh\mu$, $\Omega_{-\mu_*}=|\Omega|$, and for $\ep_0<0$,  $\Omega_{\mu}$ is the one with the minimum.
The simple  approximations for $\Omega_{\pm\mu}$ are also inferred from Fig.~\ref{frabi0}. For $\ep_0>0$, we have
\be
\begin{split}
	\Omega_{\mu}\approx\vD_{\mu}~\text{for~all}~\mu,&~
	\Omega_{-\mu}\approx\vD_{-\mu}~\text{for}~\mu<\wh\mu, \\
	\Omega_{-\wh\mu}\approx |\Omega|~\text{for}~\mu=\wh\mu,&~
	\Omega_{-\mu}\approx -\vD_{-\mu}~\text{for}~\mu>\wh\mu.
\end{split}
\lab{t0}
\ee  
The same approximations for $\ep_0<0$ require the $\Omega_{\pm\mu}\to\Omega_{\mp\mu}$ swap in every part of Eq.~\bref{t0}, see Fig.~\ref{frabi0}(b).

\section{Dressed states}\lab{sec5}
Equation~\bref{re1}  has an obvious and important class of solutions with the time-independent sideband powers - dressed (eigen) states. The Rabi oscillations stem from a superposition of  the dressed states.
Dressing the states, i.e., working with the superpositions between 
$\wt\psi_{\mu\f}$ and $\wt\psi_{\mu\s}$, rather than with the modes of the linear resonator, allows to develop a theory embracing the cases with  the arbitrary (small, large or near one) ratios of the $|\wt\psi_{\mu\s}|^2$ and $|\wt\psi_{\mu\f}|^2$ powers~\cite{prr}.

After the re-scaling, $\wt\psi_{\pm\mu\f}=e^{\pm i\phi_\f}\tilde a_{\pm\mu\f}/\sqrt{2}$, 
$\wt\psi_{\pm\mu\s}=e^{\pm i\phi_\s}\tilde a_{\pm\mu\s}/\sqrt{\gamma_{2\f}/\gamma_{2\s}}$, $\phi_\zeta=\arg\psi_{0\zeta}$,  Eq.~\bref{kk} become, 
\be
i\p_t\ket{\tilde a_\mu}=\big(\wh H_\mu+\wh V\big)\ket{\tilde a_\mu}.
\lab{master}
\ee
Here  $\ket{\tilde a_\mu}=(\tilde a_{\mu\f},\tilde a_{\mu\s},\tilde a_{-\mu\f},\tilde a_{-\mu\s})^T$ is
the state vector, 
\begin{align}
	\wh H_\mu=\left[\begin{array}{cccc}
		\vD_{\mu\f} 
		&-\tfrac{1}{2}|\Omega| e^{-i\phi}
		& 0
		&0
		\\
		-\tfrac{1}{2}|\Omega| e^{i\phi}
		& \vD_{\mu\s} 
		& 0 
		& 0
		\\
		0
		& 0 
		& -\vD_{-\mu\f} 
		&\tfrac{1}{2}|\Omega| e^{i\phi}
		\\
		0 
		& 0
		& \tfrac{1}{2}|\Omega| e^{-i\phi} 
		&-\vD_{-\mu\s}
	\end{array}\right],
	\lab{ham}
\end{align}
$\phi=2\phi_{\f}-\phi_{\s}$, and 
\begin{align}
	\wh V=\left[\begin{array}{cccc}
		-i\tfrac{1}{2}\kappa_\f
		&0
		& -\dfrac{|\Omega|^2e^{-i\phi}}{|\Omega_\s|} 
		&0
		\\
		0
		&  -i\tfrac{1}{2}\kappa_\s
		& 0 
		& 0
		\\
		\dfrac{|\Omega|^2e^{i\phi}}{|\Omega_\s|} 
		& 0 
		& -i\tfrac{1}{2}\kappa_\f
		&0
		\\
		0 
		&0
		& 0 
		&-i\tfrac{1}{2}\kappa_\s
	\end{array}\right].
	\lab{v}
\end{align}

\begin{figure}[t]
	\centering
	\includegraphics[width=0.48\textwidth]{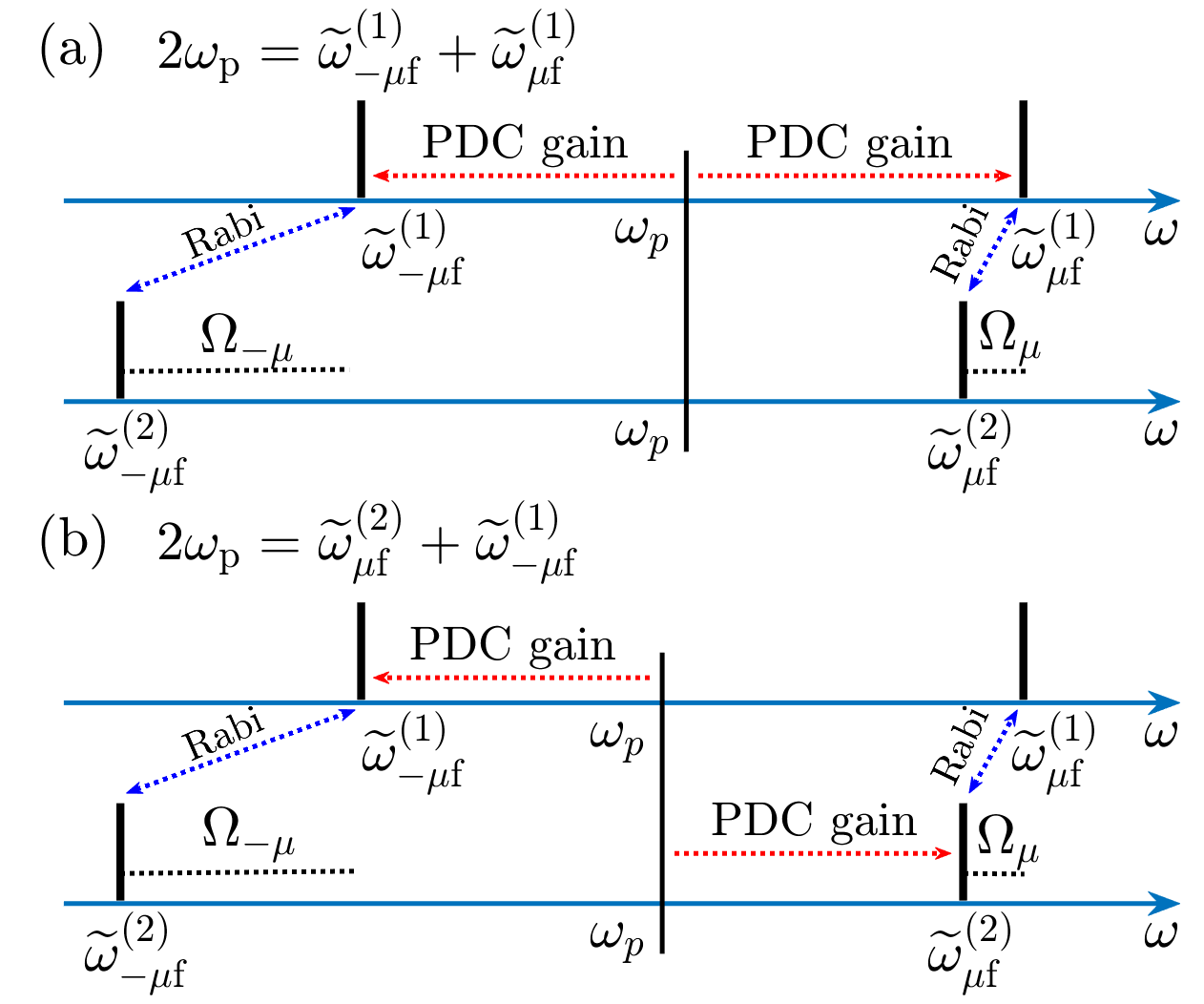}
	\caption{Diagram illustrating the dressed spectra. (a)~shows the pump frequency tuned to satisfy the intra-branch PDC condition,
		$\hbar 2\om_p=\hbar\wt\om_{\mu\f}^{(1)}+\hbar\wt\om_{-\mu\f}^{(1)}$. (b)~shows how the cross-branch PDC condition is satisfied, $\hbar 2\om_p=\hbar\wt\om_{\mu\f}^{(2)}+\hbar\wt\om_{-\mu\f}^{(1)}$.
		The red and blue arrows show how the PDC gain and Rabi flops redistribute power within the spectrum.
		The black doted lines show the Rabi frequencies, $\Omega_{\mu}$, $\Omega_{-\mu}$, and highlight their inequality.
	}
	\lab{fpdc}
\end{figure}

Setting
\begin{align}
	& \ket{\tilde a(t)}=\ket{a_\mu}\exp\{t\lambda_\mu-it\beta_\mu \},\\
	& \lambda_\mu\in\mathbb{R},~\beta_\mu \in\mathbb{R},\nn
\end{align} 
we find
\be
\left(\beta_\mu+i\lambda_\mu\right)\ket{a_\mu}
=\left(\wh H_\mu+\wh V\right)\ket{a_\mu},
\lab{master0}
\ee
where   
$\beta_\mu$ is the frequency shift, and $\lambda_\mu $ is the sideband growth rate, such that
$\lambda_\mu$ transiting from negative to positive signals 
instability of the cw-state relative to the excitation of  the $\pm\mu$  pair.

In the SC regime, see Eq.~\bref{sc0}, $\wh V$ is a perturbation 
to $\wh H_\mu$, and therefore, before incorporating $\wh V$, 
we look into the details of the dressed states,
\be
\wh H_\mu\ket{b_\mu^{(j)}}=\beta_\mu^{(j)}\ket{b_\mu^{(j)}},~j=1,2,3,4.
\lab{master1}
\ee
The eigenfrequencies, $\beta_\mu^{(j)}$, and state vectors, $\ket{b_\mu^{(j)}}$, of the
four branches of the dressed spectrum are~\cite{prr}
\be
\begin{split}
\beta_\mu^{(1)}&=\dfrac{1}{2}\big(\vD_{\mu \f}+\vD_{\mu \s}\big)+\dfrac{1}{2}\Omega_\mu,
\\
&\ket{b_\mu^{(1)}}=|\Omega|e^{-i\phi} \ket{1}+\big(\vD_{\mu}-\Omega_\mu\big)\ket{2};
\\ 
 \beta_\mu^{(2)}&=\dfrac{1}{2}\big(\vD_{\mu \f}+\vD_{\mu\s}\big)-\dfrac{1}{2}\Omega_\mu,		
\\
 &\ket{b_\mu^{(2)}}=
|\Omega|e^{-i\phi} \ket{1}+ \big(\vD_{\mu}+\Omega_\mu\big)\ket{2};
\\
\beta_\mu^{(3)}&=-\dfrac{1}{2}\big(\vD_{-\mu \f}+\vD_{-\mu \s}\big)-\dfrac{1}{2}\Omega_{-\mu},
\\
&\ket{b_\mu^{(3)}}=|\Omega|e^{i\phi} \ket{3}+\big(\vD_{-\mu}-\Omega_{-\mu}\big)\ket{4};
\\ 
\beta_\mu^{(4)}&=-\dfrac{1}{2}\big(\vD_{-\mu \f}+\vD_{-\mu \s}\big)+\dfrac{1}{2}\Omega_{-\mu},
\\
&\ket{b_\mu^{(4)}}=|\Omega|e^{i\phi} \ket{3}+\big(\vD_{-\mu}+\Omega_{-\mu}\big)\ket{4}.
\end{split}
\lab{state}
\ee

The branches $(1)$ and $(2)$ describe the Rabi induced coupling between the $+\mu$ sidebands in the fundamental and 2nd harmonic, and $(3)$, $(4)$ do the same for the $-\mu$ sidebands. 
The corresponding dressed frequencies are 
\be
\begin{split}
	&\wt\om_{\mu\f}^{(1),(2)}=\om_p+\mu D_{1}+\beta_\mu^{(1),(2)},\\
	&\wt\om_{\mu\f}^{(3),(4)}=\om_p-\mu D_{1}-\beta_\mu^{(3),(4)},\\
	&\wt\om_{\mu\s}^{(1),(2)}=2\om_p+\mu D_{1}+\beta_\mu^{(1),(2)},\\
	&\wt\om_{\mu\s}^{(3),(4)}=2\om_p-\mu D_{1}-\beta_\mu^{(3),(4)}.
\end{split}
\lab{ph0}
\ee
Taking the explicit expressions for $\vD_\mu$ and $(\vD_{\mu\f}+\vD_{\mu\s})/2$ inside $\beta_\mu^{(j)}$ one would find that $\wt\om_{\mu\zeta}^{(j)}$ are the reference frame, i.e., $D_1$, independent.

Power distribution between the fundamental and 2nd harmonic sidebands within a given branch and for a given $\mu$, is determined by 
the dressing  parameters, $\vD_{\pm\mu}^2/|\Omega|^2$. 
If the Rabi detuning is relatively large, i.e., a particular mode  is far from being the sum-frequency matched, see the points away from the zero line in Fig.~\ref{fcross}(a), then  the corresponding dressed state,
$\ket{b_\mu^{(j)}}$, tends towards an
eigenstate of the bare, i.e., $\Omega=0$, resonator. The bare states are
\be
\begin{split}
	&\ket{1}=(1,0,0,0)^T,~e^{i(M+\mu)\vta-it\om_{\mu\f}},
	\\
	& \ket{2}=(0,1,0,0)^T,~e^{i(2M+\mu)\vta-it\om_{\mu\s}},
	\\
	&\ket{3}=(0,0,1,0)^T,~e^{i(M-\mu)\vta-it\om_{-\mu\f}}
	\\ 
	& \ket{4}=(0,0,0,1)^T,~e^{i(2M-\mu)\vta-it\om_{-\mu\s}},
\end{split} 
\lab{r0}
\ee
where, for the sake of clarity, we explicitly associated each of the state vectors
to the corresponding resonator mode. 

The maximal dressing condition, $\vD_{\pm\mu}=0$, involves frequencies of the bare resonator,
while the matching points are replaced  by the avoided crossings in the dressed resonator, see  Fig.~\ref{fcross}(b). The avoided crossing between  $\wt\om_{\pm\mu\zeta}^{(1)}$ and $\wt\om_{\pm\mu\zeta}^{(2)}$ exist for every $\mu$, however,
most of them do not come to the practical, tens of MHz, 
proximity of the~$\delta=0$, apart from the ones nearest to~$\pm\mu_*$.

Noting the  symmetries 
\be
\begin{split}
	&\beta_{\mu}^{(3)}=-\beta_{-\mu}^{(1)},~\text{i.e.,}~\wt\om_{-\mu\zeta}^{(1)}=\wt\om_{\mu\zeta}^{(3)},\\
	&\beta_{\mu}^{(4)}=-\beta_{-\mu}^{(2)},~\text{i.e.,}~\wt\om_{-\mu\zeta}^{(2)}=\wt\om_{\mu\zeta}^{(4
		)},
\end{split}
\lab{ph01}
\ee
we conclude that there are two ways to proceed from this point. First, the problem could  be formulated
using the four dressed frequencies, e.g., 
$\wt\om_{\mu\f}^{(1)}$, $\wt\om_{\mu\f}^{(2)}$, 
$\wt\om_{\mu\f}^{(3)}$, $\wt\om_{\mu\f}^{(4)}$~\cite{prr}. Second, one could switch to using the two-branch formulation
and deal with $\wt\om_{\mu\f}^{(1)}$, $\wt\om_{-\mu\f}^{(1)}$, and  $\wt\om_{\mu\f}^{(2)}$,
$\wt\om_{-\mu\f}^{(2)}$. 
The latter approach is slightly more intuitive  and we choose to
follow it  here. In either case,  the four frequencies and eigenstates have to be traced. The $\wt\om_{\pm\mu\f}^{(1)}$,  $\wt\om_{\pm\mu\f}^{(2)}$ frequencies in the dressed spectrum are illustrated in Fig.~\ref{fpdc}. The $\wt\om_{\pm\mu\s}^{(1)}$,  $\wt\om_{\pm\mu\s}^{(2)}$ spectra characterize the same dressed states as $\wt\om_{\pm\mu\f}^{(1)}$,  $\wt\om_{\pm\mu\f}^{(2)}$. If the former are plotted then they would make the same spectrum as in Fig.~\ref{fpdc} apart from being centred at $2\om_p$, see Eq.~\bref{ph0}.

\begin{figure}[t]
	\centering
	\includegraphics[width=0.49\textwidth]{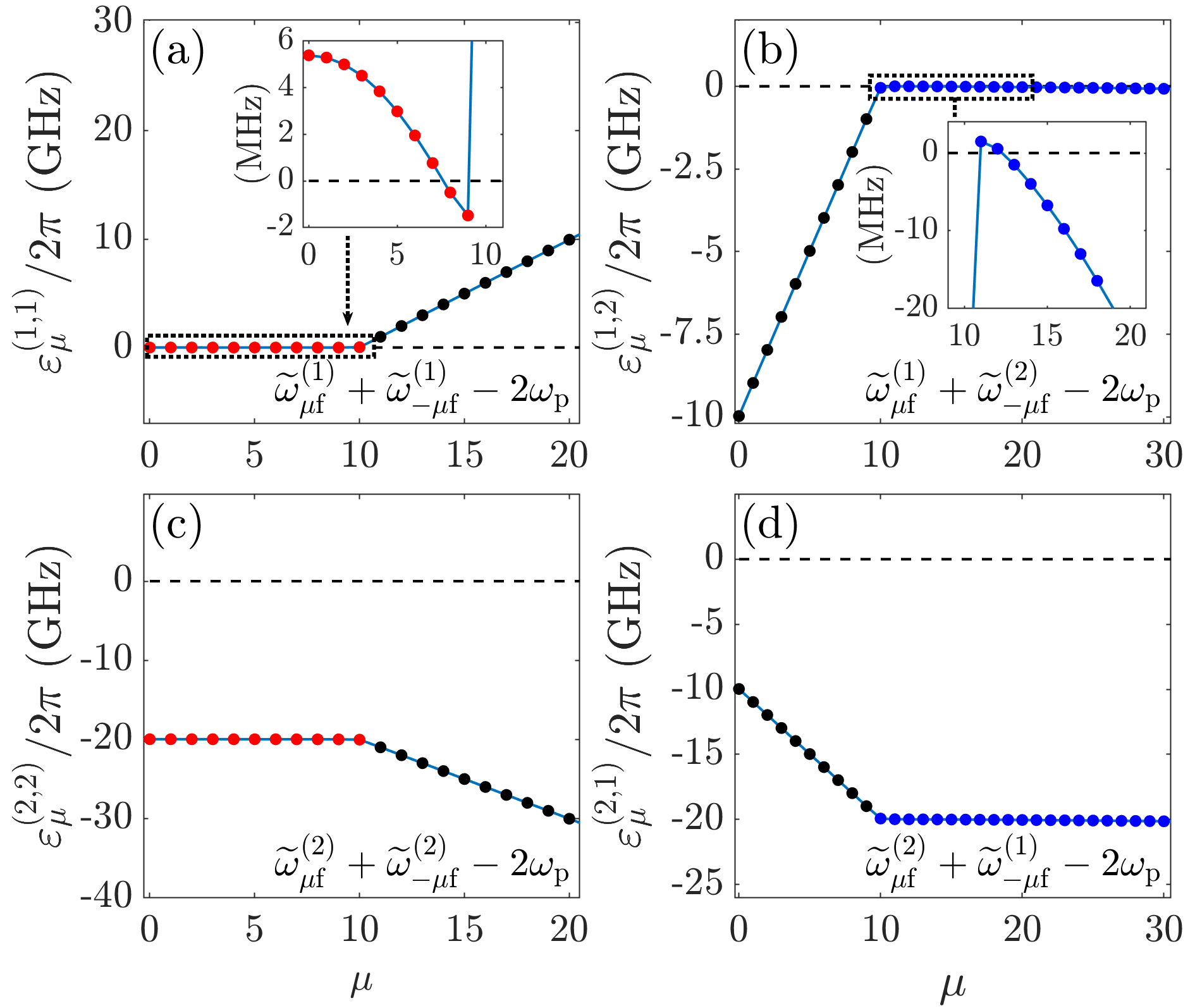}
	\caption{ Frequency matching parameters $\ep_{\mu}^{(j_1j_2)}=\wt\om_{\mu_1\f}^{(j_1)}+\wt\om_{\mu_2\f}^{(j_2)}-2\om_p$ for the four types of the parametric-down conversion (PDC) conditions.    The sideband numbers grouped around the dashed horizontal  lines, $\ep_{\mu}^{(j_1j_2)}=0$, correspond to the MHz  mismatches that can be compensated by the nonlinear effects and lead to the exact PDC frequency matching, see insets. 
		(a), (b) $\delta=2.55\kappa_\f$, $|\Omega|/2\pi=76$MHz, 
		(c), (d) $\delta=8.1\kappa_\f$, $|\Omega|/2\pi=105$MHz, 
		and $\ep_0/2\pi=10$GHz.  $\mu\approx 10$, see the corner points, corresponds to the  sum-frequency matching, $\vD_{-\mu}\approx 0$. }
	\lab{fpair}
\end{figure}

\begin{figure}[t]
	\centering
	\includegraphics[width=0.49\textwidth]{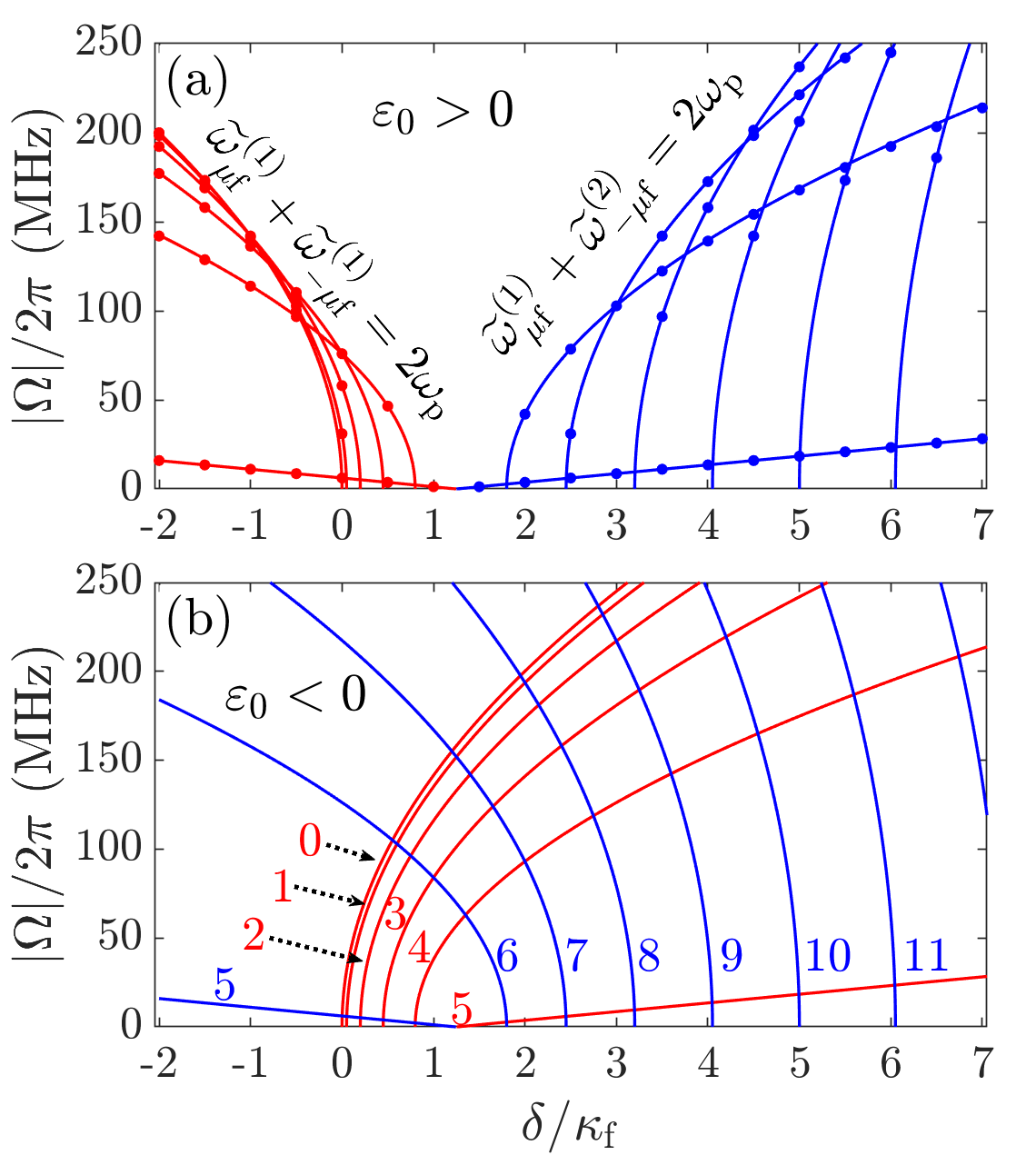}
	\caption{Lines corresponding to the exact PDC frequency matching conditions in the parameter space
		of the pump detuning, $\delta$, and of the Rabi frequency, $|\Omega|$, characterising the intra-resonator cw-power, see Fig.~\ref{fcw}. The red lines (numbered 0 to 5) correspond to  the intra-branch PDCs, which are
		 $\hbar\wt\om_{\mu\f}^{(1)}+\hbar\wt\om_{-\mu\f}^{(1)}=\hbar 2\om_p$ in (a) where $\ep_0/2\pi=5$GHz,
		 and 		 $\hbar\wt\om_{\mu\f}^{(2)}+\hbar\wt\om_{-\mu\f}^{(2)}=\hbar 2\om_p$ in (b) where $\ep_0/2\pi=-5$GHz, $0\le \mu\le 5$. The blue lines (numbered 5 to 11) correspond to  the cross-branch PDCs, which are
		 $\hbar\wt\om_{\mu\f}^{(1)}+\hbar\wt\om_{-\mu\f}^{(2)}=\hbar 2\om_p$ in (a), and
 		 $\hbar\wt\om_{\mu\f}^{(2)}+\hbar\wt\om_{-\mu\f}^{(1)}=\hbar 2\om_p$ in (b), $\mu\ge 5$. 
 		 The lines in (a) and (b) use  Eq. \eqref{syn1}, while the dots in (a) are derived from the approximate Eq.~\bref{kerr0} for $\mu\ne 5$  and Eq.~\eqref{po} for $\mu=5$. 
	}
	\lab{fbetamu}
\end{figure}

\begin{figure*}[t]
	\centering
	\includegraphics[width=1.\textwidth]{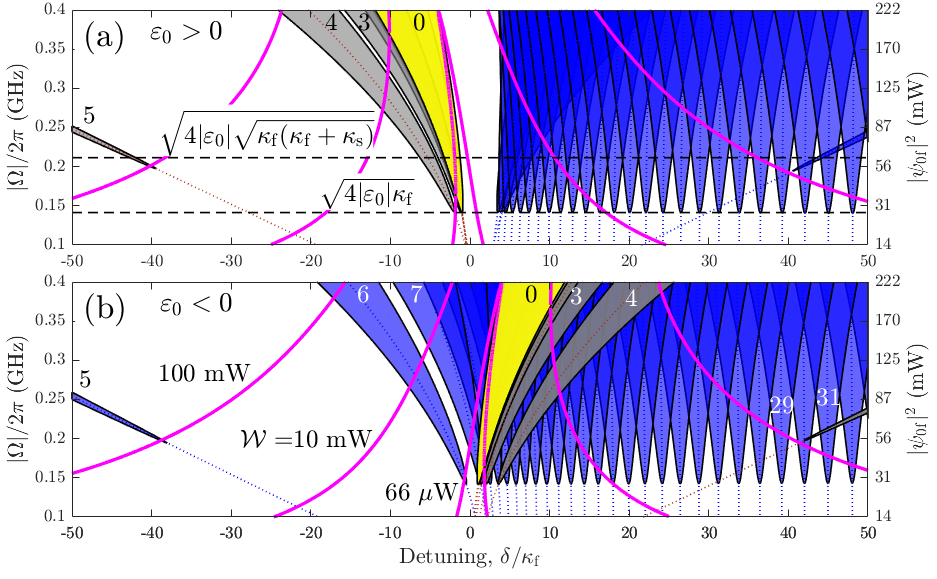}
	\caption{Parametric down-conversion (PDC) instability tongues mapped onto the parameter space span by the laser detuning, $\delta$, and the Rabi frequency, $|\Omega|$ (see the left axis), or, and equivalently, by the intra-resonator cw power, $|\psi_{0\f}|^2$ (see the right axis). 
 (a)~is for $\ep_0/2\pi=5$GHz, and (b)~is for $\ep_0/2\pi=-5$GHz. The grey shaded tongues correspond to the intra-branch PDC conditions marked by the red lines in  Fig.~\ref{fbetamu}, and the blue tongues - to the cross-branch PDCs. Some of the tongues are marked with the respective $\mu$'s, see Fig.~\ref{fbetamu} for the complete illustration of the $\mu$ ordering. The magenta lines show  $|\Omega|=|\psi_{0\f}|\sqrt{8\gamma_\f\gamma_\s}$ vs $\delta$ achieved for the laser powers $\cW=66\mu$W, $10$mW, and $100$mW. 
	}
	\lab{ftong}
\end{figure*}

\section{Dressed spectrum and  energy conservation in parametric-down conversion  (PDC)}\lab{sec6}
In any parametric system, the parametric resonance is achieved for, usually, a sequence of 
the resonance values of the drive frequency~\cite{jor}. In optical resonators in general, and in the dressed $\chi^{(2)}$ system, in particular,  this is done by tuning $\om_p$ to the mid-point between the desired sidebands, e.g.,  $\wt\om_{\mu\f}^{(1)}$ and $\wt\om_{-\mu\f}^{(1)}$. One peculiar feature of our case is that the dressed resonances depend on the pump power and frequency. Another is that for the two pairs of frequencies, $\wt\om_{\pm\mu\f}^{(1)}$, $\wt\om_{\pm\mu\f}^{(2)}$, there could be four different mid-points for the same $\mu$, and hence 
four conditions providing the maximum of the parametric gain~\cite{prr,prrcom},
\bsub
\lab{w0}
\begin{align}
	&2\hbar\om_p=\hbar\wt\om_{\mu\f}^{(1)}+\hbar\wt\om_{-\mu\f}^{(1)},
	~(\beta_\mu^{(1)}=-\beta_{-\mu}^{(1)}),
	\label{w0a}\\
	&2\hbar\om_p=\hbar\wt\om_{\mu\f}^{(2)}+\hbar\wt\om_{-\mu\f}^{(2)}, 
	~(\beta_\mu^{(2)}=-\beta_{-\mu}^{(2)}),
	\label{w0b}\\
	& 2\hbar\om_p=\hbar\wt\om_{\mu\f}^{(1)}+\hbar\wt\om_{-\mu\f}^{(2)},
    ~(\beta_\mu^{(1)}=-\beta_{-\mu}^{(2)}),
	\label{w0c}\\
	&2\hbar\om_p=\hbar\wt\om_{\mu\f}^{(2)}+\hbar\wt\om_{-\mu\f}^{(1)},
	~(\beta_\mu^{(2)}=-\beta_{-\mu}^{(1)}).
	\label{w0d}  
\end{align}
\esub
 
The auxiliary frequency $\wt\om_\mu=\om_p+\mu D_{1\f}+\tfrac{1}{2}(\vD_{\mu\f}+\vD_{\mu\s})$ 
makes the role of the Rabi splitting in the above more transparent,
\bsub
\lab{z0}
\begin{align}
	&2\hbar\om_p=\hbar(\wt\om_{\mu}+\tfrac{1}{2}\Omega_{\mu})+\hbar(\wt\om_{-\mu}
	+\tfrac{1}{2}\Omega_{-\mu}),
	\label{z0a}\\
	&2\hbar\om_p=\hbar(\wt\om_{\mu}-\tfrac{1}{2}\Omega_{\mu})+\hbar(\wt\om_{-\mu}
	-\tfrac{1}{2}\Omega_{-\mu}),
	\label{z0b}\\
	& 2\hbar\om_p=\hbar(\wt\om_{\mu}+\tfrac{1}{2}\Omega_{\mu})+\hbar(\wt\om_{-\mu}
	-\tfrac{1}{2}\Omega_{-\mu}),
	\label{z0c}\\
	&2\hbar\om_p=\hbar(\wt\om_{\mu}-\tfrac{1}{2}\Omega_{\mu})+\hbar(\wt\om_{-\mu}
	+\tfrac{1}{2}\Omega_{-\mu}).
	\label{z0d}  
\end{align}
\esub

The first pair of conditions, Eqs.~\bref{w0a}, \bref{w0b}, corresponds to the intra-branch  PDC, which is satisfied  by tuning the pump frequency to the mid-point between the $\wt\om_{\mu\f}^{(j)}$ and $\wt\om_{-\mu\f}^{(j)}$.
The second pair, Eqs.~\bref{w0c}, \bref{w0d}, are the cross-branch PDC conditions. They are satisfied by $\om_p$ being tuned to the mid-point between the  $\mu$ and $-\mu$ sidebands from the two different branches of the dressed spectrum. 

Figure~\ref{fpdc} illustrates achieving frequency matching for 
the intra- and cross-branch cases,  and also shows how 
the effective Rabi frequencies, $\Omega_{\pm\mu}$, come into play.
$\Omega_\mu$ and $\Omega_{-\mu}$ are generally very different, and coincide only for $\mu=0$, see Fig.~\ref{frabi0}.

To gain further important insights into the PDC conditions in Eq.~\bref{w0},  we  rearrange them   as 
\be
\ep_\mu^{(j_1,j_2)}=\wt\om_{\mu\f}^{(j_1)}+\wt\om_{-\mu\f}^{(j_2)}-2\om_p
\ee
and  plot $\ep_\mu^{(j_1,j_2)}$ vs $\mu$, see Fig.~\ref{fpair}.
The sideband numbers grouped around the zero lines  correspond to the MHz level mismatches 
that are more easily compensated by the nonlinear effects providing $\ep_\mu^{(j_1,j_2)}=0$. 
The sidebands with $\ep_{\mu}^{(j_1,j_2)}$ detuned 
away from the zero by the GHz offsets (see the black dots in Figs.~\ref{fpair}(a), (b))
are  cutting-off from the groups of the PDC capable mode numbers. 
The cut-off is happening for $\mu$'s around  $\mu_*$,  
corresponding to the sum-frequency matching, see Eq.~\bref{epe}.

Figure \ref{fpair}(a) shows that the intra-branch condition in Eq.~\bref{w0a} can be satisfied for  
$0\le\mu\lesssim \mu_*$, while $\ep_0$ is set to be positive. The second intra-branch condition, Eq.~\bref{w0b},  see Fig.~\ref{fpair}(c), is shifted away from zero by $\approx 2\ep_0$, and will be swapped with the first one for $\ep_0\to-\ep_0$. Figures \ref{fpair}(b) and \ref{fpair}(d) show the cross-branch PDC conditions, with one of them being  satisfied for
$\mu\gtrsim \mu_*$. 

The challenge with resolving  Eq.~\bref{w0} analytically for either $\Omega$ or $\delta$ is in the occurrences of them under the square root sign in the equation for $\Omega_\mu$, see Table~\ref{t2}. However, the algebra is proceedable~\cite{prr,prrcom}, and leads to finding that all four PDC conditions are resolved by $|\Omega|=|\Omega_{\text{pdc}}|$, where
\begin{align}
	|\Omega_{\text{pdc}}|^2&=4
	(\vD_{\mu \f}+\vD_{-\mu \f}) 
	(\vD_{\mu \s}+\vD_{-\mu \s})
	\nn
	\\
	&\times
	\frac{(\vD_{\mu \f}+\vD_{-\mu \s})
		(\vD_{\mu \s}+\vD_{-\mu \f})
	}
	{(
		\vD_{\mu \f}+\vD_{-\mu \f}+
		\vD_{\mu \s}+\vD_{-\mu \s})^2}.
	\lab{syn1}
\end{align}

Plots of $|\Omega_{\text{pdc}}|$ vs $\delta$ for positive and negative $\ep_0$, and their associations with the PDC conditions are shown in Fig.~\ref{fbetamu}. 
These plots in Fig.~\ref{fbetamu} could be compared with the temperature tuning diagrams of the parametric oscillators, see, e.g., \cite{byer0,byer1,st0}, but in our case, the temperature is assumed fixed, while the tuning parameters are the pump power expressed via $|\Omega|$, 
and the pump frequency. The range of $|\Omega|$s considered by us provides the relatively small intraresonator powers, order of mW to $<1$W, cf., the left and right axes in Fig.~\ref{ftong}. 

If $|\ep_0|$ dominates over all $\delta$'s, and   $\mu_*$ falls  between the two nearest integers, i.e.,
the exact sum-frequency matching point has been missed, 
then Eq.~\bref{syn1}  simplifies to 
\be
|\Omega_{\text{pdc}}|^2\approx -4\ep_0\left(\delta-\delta_{\mu\f}\right)
\left[1-\frac{\mu^2}{\mu_*^2}\right],~ \delta_{\mu\f}=-\tfrac{1}{2}D_{2\f}\mu^2,
\lab{kerr0}
\ee
see Appendix~\ref{ap5} for details.

Eq.~\bref{kerr0} reveals what has been described above based on the numerical plots. First, one can see that the resonances converge to points $\delta=\delta_{\mu\f}$ for $\Omega\to 0$, which corresponds to the zero of the first bracket in the numerator of Eq.~\bref{syn1}. Second, the direction of the nonlinearity induced tilts of the resonances depends  on the sign of $\ep_0$ and 
the value of $\mu$, see Fig.~\ref{fbetamu}. If $\ep_0<0$, then the tilt is towards $\delta>\delta_{\mu\f}$ for $0\le\mu<\mu_*$,
and towards $\delta<\delta_{\mu\f}$ for $\mu>\mu_*$. $\ep_0>0$ changes the tilt direction for the two groups of modes. 

Figure~\ref{fbetamu} also shows a good agreement between the exact and approximate $|\Omega_{\text{pdc}}|^2$ vs $\delta$ dependencies. 
Analytical approximation for $|\Omega_{\text{pdc}}|$ in the case
when the sum-frequency process is either nearly or exactly matched, i.e., 
$\mu=\wh\mu\approx\mu_*$, is considered in  Appendix~\ref{ap5}.

To summarise  - the intra-branch PDC conditions 
are satisfied for a compact group of the sideband numbers, 
\be 0\le\mu\le\mu_*,~\text{i.e., for}~
\left[1-\frac{\mu^2}{\mu_*^2}\right]\ge0,
\lab{e47}
\ee
while the cross-branch ones are engaged for the unbound set of sidebands,
\be\mu\ge\mu_*,~\text{i.e., for}~
\left[1-\frac{\mu^2}{\mu_*^2}\right]\le 0.\ee 
Thus,  the sum-frequency matched sideband, $\mu_*$ or $-\mu_*$, defines the transition between the 
two different PDC scenarios.

\section{PDC instability tongues}\lab{sec7}
While the PDC frequency matching provides conditions for the maximal parametric gain, 
the latter still needs  to overcome the dissipation in order to trigger  the exponential growth of sidebands, i.e., to induce the cw instabilities. Regions of the PDC  instabilities for every $\pm\mu$ pair of sidebands can be computed numerically by solving the eigenvalue problem in Eq.~\bref{master0} and plotting the lines $\lambda_\mu=0$~\cite{prr}.
Every $\mu$-specific instability area is represented by a tongue-like domain shaped around 
the respective $|\Omega_{\text{pdc}}|$ vs $\delta$ line, cf., Figs.~\ref{fbetamu} and \ref{ftong}. 

The intra-branch instabilities, Eqs.~\bref{w0a} and \bref{w0b}, 
are coloured in grey in~Fig.~\ref{ftong}. While, the cross-branch ones, Eqs.~\bref{w0c}, \bref{w0d}, are shown in blue.
The magenta lines show $|\Omega|$'s vs $\delta$ corresponding to the cw-state achieved for three representative values of the laser power, $\cW=66\mu$W, $10$mW, and $100$mW. The cw-state is expressed via $\Omega$ as per Eq.~\bref{ws10}, while $\Omega$ itself is a solution of
\be
\Omega=\sqrt{\frac{\kappa_\f\kappa_\s}{2}}\sqrt{\frac{\cW~}{\cW_*}}
\frac{\kappa_\f}{\Omega_\f}\left[
1-\frac{|\Omega|^2}{\Omega_\f\Omega_\s}\right]^{-1},\lab{cw8}
\ee
where $\Omega_\f=\delta-i\tfrac{1}{2}\kappa_\f$, 
$\Omega_\s=8(2\delta-i\tfrac{1}{2}\kappa_\s)-8\ep_0$,  $\cW$ is the laser power in Watts, and $\cW_*$ is its scaling, see Eq.~\bref{star}.
Taking modulus squared of Eq.~\bref{cw8} we find a real cubic equation for $|\Omega|^2$, that 
can have either one or three
positive roots, with the latter case signalling the cw-bistability, see Fig.~\ref{fcw}, and further details in Appendix~\ref{ap3}. 

Thus, a scan of the laser frequency, $\om_p$,  would go along an individual power-defined  path crossing the different PDC domains, see the magenta lines in Fig.~\ref{ftong}. The yellow shading marks the $\mu=0$ tongue  embracing  the middle branch of the bistability loop, see Fig.~\ref{fcw}. For the Kerr resonators, the similar tongue diagrams  were recently reported 
in Refs.~\cite{arnold,pra}.

\begin{figure*}[t]
	\centering
	\includegraphics[width=1.\textwidth]{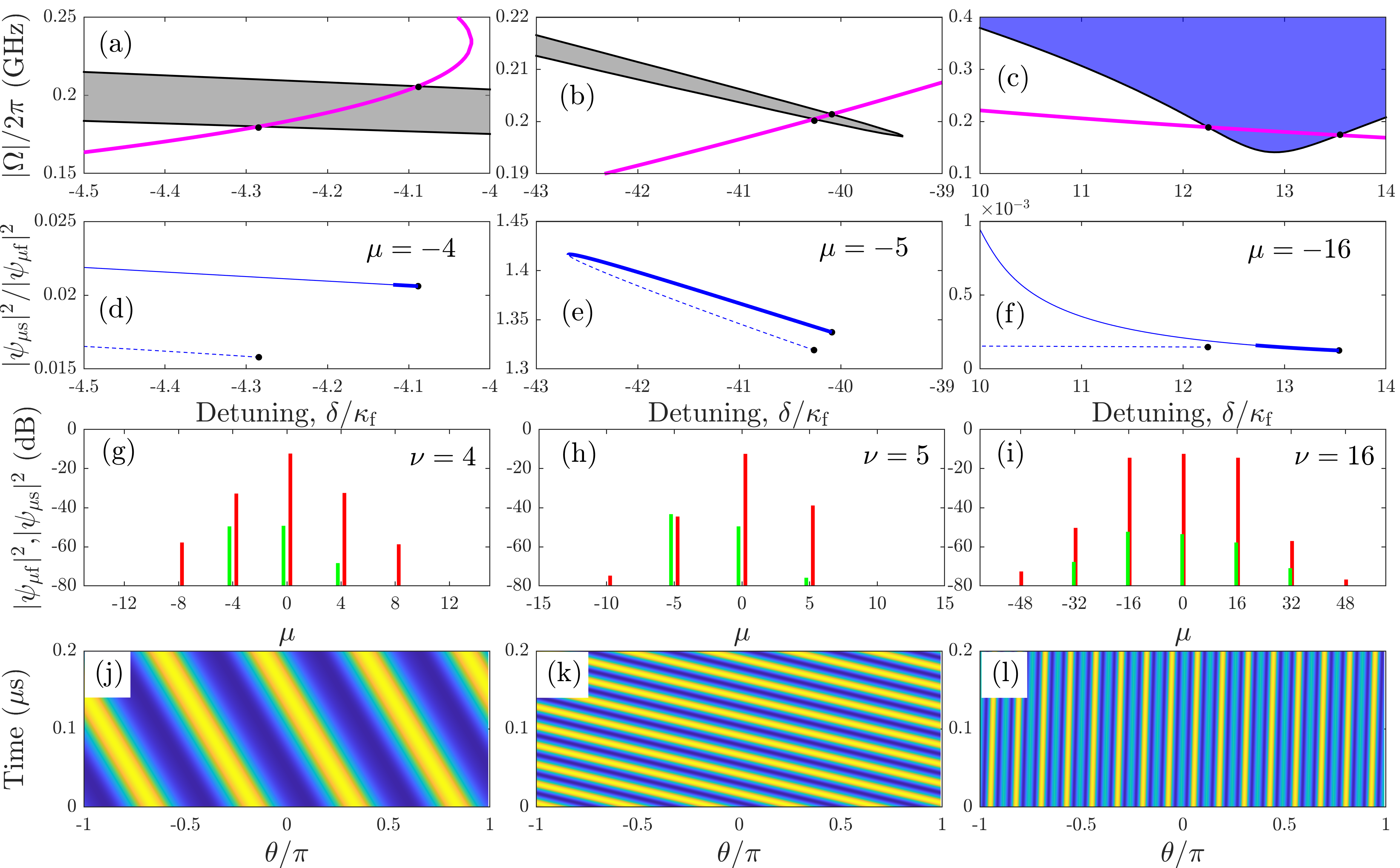}
	\caption{Examples of the Turing-pattern frequency combs for $\ep_0/2\pi=5$GHz. First and second columns show the combs emerging from the intra-branch PDC instability tongues (grey areas in (a,b)), and the third column shows the cross-branch PDC tongue (blue area in (c)) and the associated comb. Magenta lines in (a-c) show the cw-states, $|\Omega|=|\psi_{0\f}|\sqrt{8\gamma_\f\gamma_\s}$ vs $\delta$.  The respective laser powers $\cW=0.664$mW (a), $103$mW(b) and $10$mW(c). $|\Omega|/2\pi=0.2$GHz corresponds to the intraresonator power $|\psi_{0\f}|^2\simeq 55$mW. The panels (d,e,f) in the second line show the ratio of the 2nd harmonic and fundamental powers in the indicated sidebands. $\mu=-5$ case corresponds to the sum-frequency matching leading to the best conversion efficiency. The thick full lines are stable solutions, and the thin  lines (full and dashed) are unstable. (g,h,i) are the self-explanatory spectra of the Turing patterns (red-fundamental, green- 2nd harmonic). (j,k,l) are the space-time profiles of the corresponding Turing-patterns. The~tilt relative to the vertical axis characterizes deviation of the pattern repetition rate from $D_{1\f}$. 
	}
	\lab{ftp}
\end{figure*}

\begin{figure*}[t]
	\centering
	\includegraphics[width=1.\textwidth]{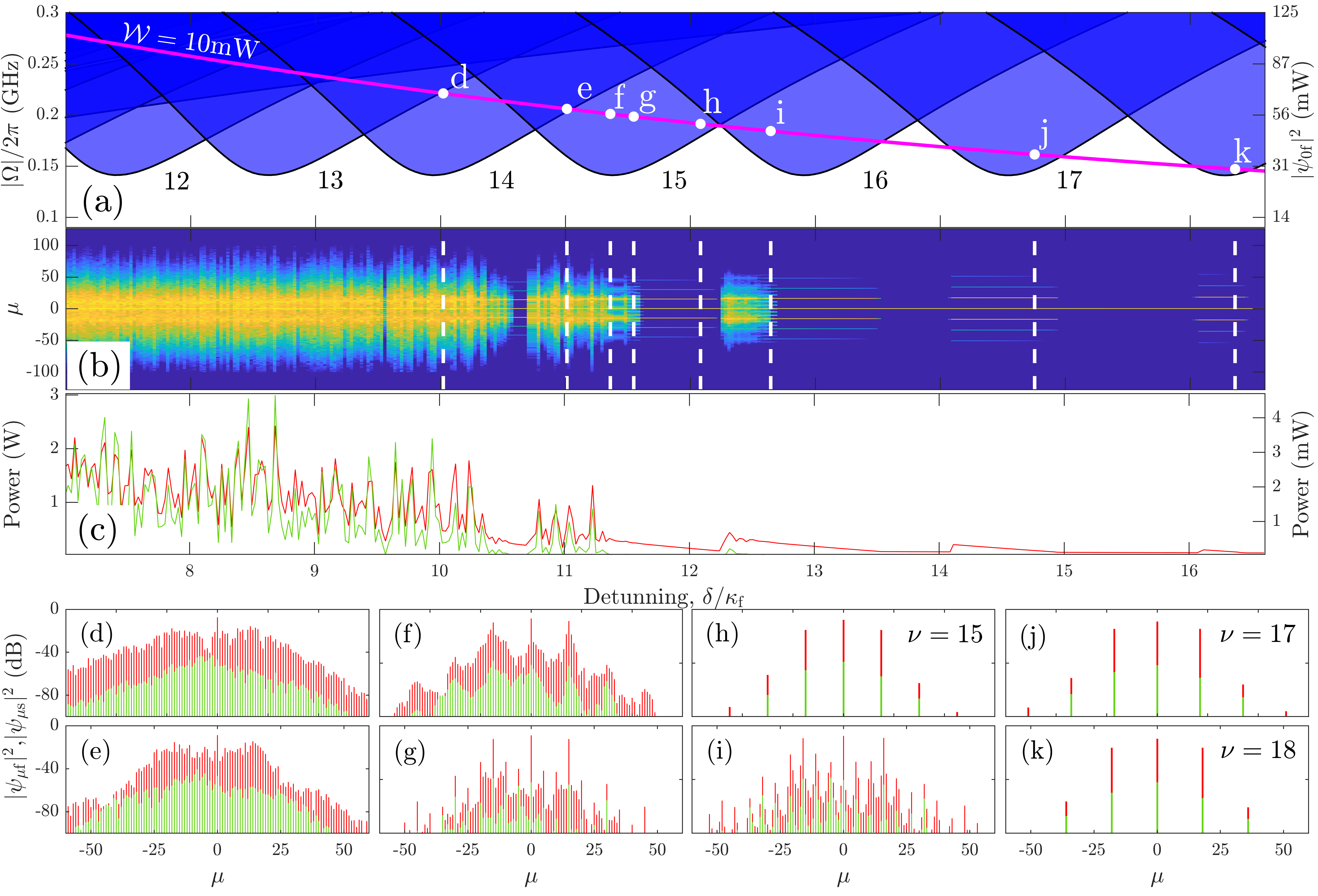}
	\caption{(a) PDC instability tongues for $\mu=12-18$,  $\ep_0/2\pi=5$GHz as in Fig.~\ref{ftong}(a). The
		magenta line shows  the cw-state for the laser power $\cW=10$mW. The left axes
		is the Rabi frequency, $|\Omega|$, and the right axis is the intra-resonator cw power, $|\psi_{0\f}|^2$.
		(b) The spectrum computed from the scan of the cw-frequency along the magenta line in (a).
		(c) The integrated intra-resonator powers corresponding to (b). 
		The red line (left axis) is the fundamental, and the green line (right axis) is the second harmonic.
		Panels in the two bottom lines show the comb spectra computed at the detunings marked with the respective letters in (a), and with the dashed white lines in (b).  In the two bottom rows, the shorter bars are plotted in green and the longer ones in red.
	}
	\lab{tur1}
\end{figure*}

\begin{figure*}[t]
\centering
\includegraphics[width=1.\textwidth]{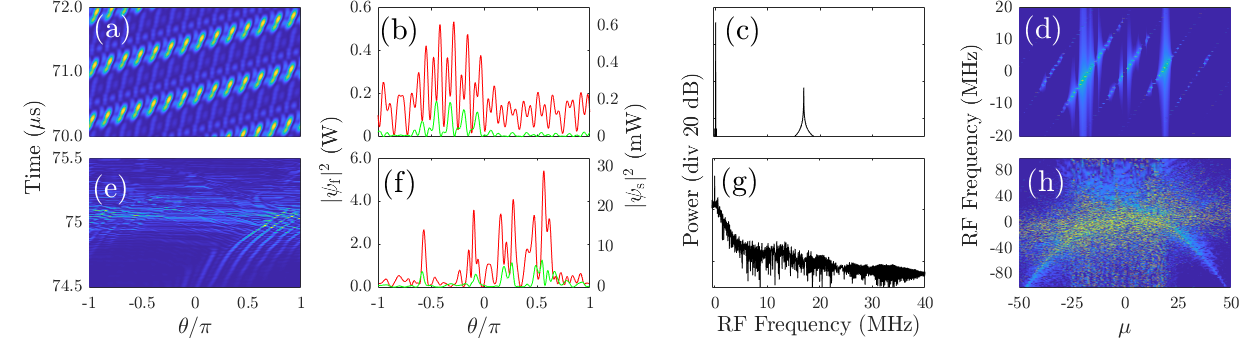}
\caption{The top row shows the breather state corresponding to the point 'f' in Fig.~\ref{tur1}(a). The bottom row is the turbulent state at the point 'e' in Fig.~\ref{tur1}(a). The 1st and 2nd columns compare the spatio-temporal dynamics (fundamental) and snapshots of the spatial profiles. The 3rd and 4th columns show the RF spectra of net powers (3rd), $|FFT\sum_\mu|\psi_{\mu\f}(t)|^2|$, and the per-mode RF spectra (4th), 
$|FFT\{\psi_{\mu\f}(t)\}|^2$. One can see that the breather state consists from the five coherent sub-combs with the different offsets and the same repetition rates, while the turbulent state shows no modelocking signs.}
	\lab{tur2}
\end{figure*}

\section{Parametric thresholds}\lab{sec8}
PDC thresholds, i.e., the minimal intra-resonator powers triggering 
the exponential growth of the $\pm\mu$ sideband pairs,  
happen at the tips of the instability tongues, see Fig.~\ref{ftong}.
To find the threshold when the system is confined to the $|\mu|$-specific PDC lines, see~Figs.~\ref{fbetamu},~\ref{ftong},
we apply the degenerate state  perturbation theory to Eq.~\bref{master0}  by treating  $\wh V$ as a perturbation 
to $\wh H_\mu$, which is valid in the SC regime. 
The generic condition for the parametric gain to overcome losses has been introduced in Ref.~\cite{prr},
\be
V_\mu^{(j_1j_2)}\cdot V_\mu^{(j_2j_1)}
=V_\mu^{(j_1j_1)}\cdot V_\mu^{(j_2j_2)},
\lab{thr0}
\ee 
where  $V_\mu^{(j_1j_2)}=\bra{b_\mu^{(j_1)}}\wh V\ket{b_\mu^{(j_2)}}$
are the matrix elements of~$\wh V$. 

Opening up Eq.~\bref{thr0} for $(j_1,j_2)=(1,3)$, 
$(j_1,j_2)=(2,4)$, $(j_1,j_2)=(1,4)$, and $(j_1,j_2)=(2,3)$  
yields  four threshold conditions,
\bsub
\lab{th}
\begin{align}
	&\left[\kappa_\f+\kappa_\s\frac{(\vD_\mu-\Omega_\mu)^2}{|\Omega|^2}\right]
	\left[\kappa_\f+\kappa_\s\frac{(\vD_{-\mu}-\Omega_{-\mu})^2}{|\Omega|^2}\right]
	=\frac{4|\Omega|^4}{|\Omega_\s|^2},
	\lab{tha}
	\\
	&\left[\kappa_\f+\kappa_\s\frac{(\vD_\mu+\Omega_\mu)^2}{|\Omega|^2}\right]
	\left[\kappa_\f+\kappa_\s\frac{(\vD_{-\mu}+\Omega_{-\mu})^2}{|\Omega|^2}\right]
	=\frac{4|\Omega|^4}{|\Omega_\s|^2},
	\lab{thb}\\
	&\left[\kappa_\f+\kappa_\s\frac{(\vD_\mu-\Omega_\mu)^2}{|\Omega|^2}\right]
	\left[\kappa_\f+\kappa_\s\frac{(\vD_{-\mu}+\Omega_{-\mu})^2}{|\Omega|^2}\right]
	=\frac{4|\Omega|^4}{|\Omega_\s|^2},
	\lab{thc}
	\\
	&\left[\kappa_\f+\kappa_\s\frac{(\vD_\mu+\Omega_\mu)^2}{|\Omega|^2}\right]
	\left[\kappa_\f+\kappa_\s\frac{(\vD_{-\mu}-\Omega_{-\mu})^2}{|\Omega|^2}\right]
	=\frac{4|\Omega|^4}{|\Omega_\s|^2}. 
	\lab{thd}
\end{align}
\esub
Eqs.~\bref{th} express the balance between the PDC gain (right) and the net loss (left).
The  2nd harmonic losses, $\kappa_\s$ are weighted by the coefficients characterising the power distribution between the components in the state vectors, see~Eq.~\bref{state}. 
The explicit threshold condition presented in Ref.~\cite{prr} 
transforms to Eq.~\bref{thc} after making use of 
the identity $\frac{|\Omega|}{\Omega_\mu-\vD_\mu}
=\frac{\Omega_\mu+\vD_\mu}{|\Omega|}$.

We now note that in the SC regime $|\Omega_\s|\approx 8|\ep_0|$, see Eq.~\bref{pa2}, and hence the right-hand sides in Eq.~\bref{th} are approximated with $|\Omega|^4/16|\ep_0|^2$. 
To simplify the threshold conditions we make use of the approximations in Eq.~\bref{t0} related to the case of $\ep_0>0$ and $\Omega_{-\wh\mu}\approx |\Omega|$, see Fig.~\ref{frabi0}(a).
Then, for $\mu<\wh\mu$ both coefficients after $\kappa_\s$ in Eq.~\bref{tha} 
are small and can be omitted in the leading order, 
so that the threshold is determined by $\kappa_\f$ only and is well approximated by,
\be
|\Omega_{\text{th}}^{(\mu)}|^2\approx 4|\ep_0|\kappa_\f,~\mu\ne\wh\mu,
\lab{th1}
\ee
see the grey tongs in Fig.~\ref{ftong}(a). 
For $\mu>\wh\mu$ the left-hand side of Eq.~\bref{tha} becomes $\approx 4\kappa_\f\kappa_s|\vD_{-\mu}|^2/|\Omega|^2$, which creates  
prohibitively large power thresholds, see the cut-off transitions in Fig.~\ref{fpair}. 

For $\mu=\wh\mu$, the  threshold is approximated by
\be
|\Omega_{\text{th}}^{(\wh\mu)}|^2\approx 4|\ep_0|\sqrt{\kappa_\f(\kappa_\f+\kappa_\s)}.
\lab{th2}
\ee
$\kappa_\s$ is now also impacting the threshold, but still in a way that is not equally important with $\kappa_\f$. This is because the powers of the fundamental and second harmonic are balanced only for $\wt\om_{-\wh\mu\f}^{(1)}$ and $\wt\om_{-\wh\mu\s}^{(1)}$, but not for $\wt\om_{\wh\mu\f}^{(1)}$ and $\wt\om_{\wh\mu\s}^{(1)}$ sidebands. In the latter pair, the sum-frequency condition is mismatched, and hence the 2nd harmonic sideband is still very weak and can be disregarded, i.e., $\vD_{\wh\mu}\approx\Omega_{\wh\mu}$ in
Eq.~\bref{tha}. While, the approximations that work in the minus bracket are
$\vD_{-\wh\mu}\approx 0$, and $\Omega_{-\wh\mu}\approx |\Omega|$.

The first cross-branch condition, Eq.~\bref{thc}, has practical threshold at $\mu=\wh\mu$ as in Eq.~\bref{th2},
and for $\mu>\wh\mu$ as in Eq.~\bref{th1}, see the blue tongues in Fig.~\ref{ftong}(a).
The second intra-branch, Eq.~\bref{thb}, and second cross-branch, Eq.~\bref{thd}, conditions do not create practical thresholds for $\ep_0>0$, and  play their roles for  $\ep_0<0$, see Figs.~\ref{fbetamu}(b), \ref{ftong}(b). 

The analytical estimates for the  detuning values where the $\mu$-specific instabilities 
first happen, i.e., locations of the tips of the instability 
tongues and the respective laser powers are derived in Appendix~\ref{ap6}.

\section{Envelope and coupled-mode equations for modelocked combs}\lab{sec9}
We anticipate that the frequency comb solutions bifurcate from the $\mu$-specific boundaries of the instability tongues in Fig.~\ref{ftong}, cf., Ref.~\cite{arnold}. Since we are going to continue to number the sidebands within the combs with $\mu$, we use below the letter $\nu=1,2,3,\dots$ to mark the comb states with the sideband spacing given by $\nu$.
We seek the modelocked combs as the solutions of the equation that couple all the modes through all the allowed nonlinear coupling terms, see Eq.~\bref{mm} in Appendix~\ref{ap1}. 

The modelocked  combs are assumed to have  the period $2\pi/\nu$, i.e., $\psi_\zeta(t,\vta)=\psi_\zeta(t,\vta+2\pi/\nu)$, and, therefore, we use the substitution
\be
\psi_\zeta(t,\vta)=\Psi_{\nu\zeta}(\ta_\nu),~
\ta_\nu=\nu(\vta-D_{1\nu} t).
\lab{s1}
\ee
Here $\ta_\nu$ is a new auxiliary coordinate, such that the period $2\pi$ in $\ta_\nu$ corresponds to the period $2\pi/\nu$ in $\ta$, $\Psi_{\nu\zeta}(\ta_\nu)=\Psi_{\nu\zeta}\left(\ta_\nu+2\pi\right)$. 
$D_{1\nu}$ is an unknown comb repetition rate generally different from either 
$D_{1\f}$ or $D_{1\s}$. If the reference frame is chosen to rotate with $D_{1\f}$, 
then $D_{1\nu}\ne D_{1\f}$ would imply the relative rotation with the $D_{1\nu}- D_{1\f}$ rate, leading to the tilted spatio-temporal profiles like in the bottom row in Fig.~\ref{ftp}

The selection mechanisms of the velocity of the dissipative $\chi^{(2)}$ solitons, equivalent to the selection of  $D_{1\nu}$,  have been  discussed  
before the microresonator combs  came into the existence~\cite{preold}.
This selection is a generic aspect also encountered in, e.g., 
the equations with the higher-order dispersion terms~\cite{mil}, and 
in the cases showing the spontaneous symmetry breaking effects~\cite{arnold}. 

Substituting Eq.~\bref{s1} into Eq.~\bref{mm} we find
\be
\begin{split}
&\delta\Psi_{\nu\f}-i\nu (D_{1\f}-D_{1\nu}) \frac{d\Psi_{\nu\f}}{d\ta_\nu}-\frac{\nu^2 D_{2\f}}{2}\frac{d^2\Psi_{\nu\f}}{d\ta_\nu^2}\\-& \gamma_\f\Psi_{\nu\s}\Psi_{\nu\f}^*-\frac{i\kappa_\f}{2}
\big(\Psi_{\nu\f}-\cH \big)=0,\\
&(2\delta-\ep_0)\Psi_{\nu\s}-i\nu (D_{1\s}-D_{1\nu}) \frac{d\Psi_{\nu\s}}{d\ta_\nu}- \frac{\nu^2 D_{2\s}}{2}\frac{d^2\Psi_{\nu\s}}{d\ta_\nu^2}\\-& \gamma_\s\Psi_{\nu\f}^2-\frac{i\kappa_\s}{2}
\Psi_{\nu\s}=0.
\end{split}
\lab{tp}
\ee
As one can see, the use of $\ta_\nu$ as an argument has allowed to 
conveniently sort the modelocked combs  by the spacing, $\nu$, their sidebands make in the momentum space,
since $\nu$ enters Eq.~\bref{tp} explicitly.
In fact, $\nu$ plays a role of the Bloch momentum which is now quantised, unlike the one that varies continuously in the theory of the unbound crystal lattices~\cite{kit} and resonators~\cite{gom}.

For the frequency combs with the spatial period $2\pi/\nu$, the modes making  
non-zero contributions to
the $\Psi_{\nu\zeta}$ have numbers $\mu=\nu m$, where 
$m=0, \pm 1,\pm 2,\pm 3,\dots$ is another integer,
\be
\Psi_{\nu\zeta}=\sum_{m=-\infty}^{\infty}\Psi_{m\nu\zeta } e^{i m\ta_\nu}=
\sum_{m=-\infty}^{\infty}\Psi_{m\nu\zeta } e^{i m\nu(\vta-D_{1\nu}t)}.
\lab{fur}
\ee 
Here $\Psi_{m\nu\zeta}$ are constants satisfying an algebraic system of equations, see Eq.~\bref{tp1}.
The repetition rate with which the $2\pi/\nu$ state is reproducing itself  while rotating in the resonator
is $\nu D_{1\nu}/2\pi$.

The coupling between the different  $m\nu$ sidebands is provided by the sequence, i.e., cascade, of the sum-frequency and difference-frequency events, which become evident on substituting Eq.~\bref{fur} into Eq.~\bref{tp},
\bsub
\lab{tp1}
\begin{align}
	&	\vD_{m\nu \f}\Psi_{m\nu\f} - \frac{i\kappa_\f}{2}
	\big(\Psi_{m\nu\f}-\wh\delta_{m,0}\cH\big)\nn \\
	&	-\gamma_\f\sum_{m_1 m_2}\wh\delta_{m,m_1-m_2}\Psi_{m_1\nu\s}\Psi^*_{m_2\nu\f}=0,
	\lab{tp1a}\\
	&	\vD_{m\nu \s}\Psi_{m\nu\s} - \frac{i\kappa_\s}{2}\Psi_{m\nu\s}\nn \\	
	&-\gamma_\s\sum_{m_1 m_2}\wh\delta_{m,m_1+m_2}\Psi_{m_1\nu\f}\Psi_{m_2\nu\f}=0.
	\lab{tp1b}
\end{align}
\esub
The sideband detunings, $\vD_{m\nu \zeta}$, are defined in Eq.~\bref{detu}, where $\mu=m\nu$.  $\wh\delta_{m,m_1\pm m_2}=1$  for $m=m_1\pm m_2$ and is 
zero otherwise. Hence, every term inside the nonlinear sums in Eqs.~\bref{tp1a} and 
\bref{tp1b} corresponds to the momentum conservation laws,
\bsub
\begin{align}
&\hbar (M+m\nu)=\hbar (2M+m_1\nu)-\hbar (M+m_2\nu),\lab{msa}\\
&\hbar (2M+m\nu)=\hbar (M+m_1\nu)+\hbar (M+m_2\nu),\lab{msb}
\end{align}
\lab{ms}
\esub
describing the sum- and difference-frequency cascades. 
The left- and right-hand sides of Eq.~\bref{ms} correspond to the 
linear and nonlinear terms in Eq.~\bref{tp1}, respectively.

\section{Turing-pattern combs}\lab{sec10}
The comb equations, Eq.~\bref{tp} or Eq.~\bref{tp1},  have been solved by us with a Newton method allowing 
to self-consistently find the sideband amplitudes,  $\Psi_{m\nu\zeta}$, 
and the comb repetition rate, $D_{1\nu}$. 
Figure \ref{ftp} shows how the  comb branches with different periods, $2\pi/\nu$, 
emanate from the boundaries of the respective instability tongues. 

The sparse combs, $\nu\gg 1$, described by a combination of  few noticeable $\mu=\nu m$ sidebands have been observed and modelled in connection to several recent $\chi^{(2)}$-resonator experiments, see, e.g.,~\cite{fins,jan1,jan2,mash}.
In the context of the Kerr microresonators, the combs with the sparse spectra, as in 
Fig.~\ref{ftp}, are often called - the Turing-pattern frequency combs~\cite{che,men,arnold}.
The Kerr microresonator instability tongues and their connection to the Turing-patterns 
have been reported in Refs.~\cite{arnold,pra}.
The prior to the ring microresonator era results on the spatial pattern formation in the planar 
$\chi^{(2)}$ resonators can be found in, e.g., Refs.~\cite{oppo,stal,longhi2,etr,ol1,ol2,skrd}.

A typical bifurcation scenario that we found is that the two 
Turing-comb branches split from the two edges of the tongues  and 
extend well beyond the tongues. The branch
crossing into the tongue area can be stable (full lines in the second row 
in Fig.~\ref{ftp}), while the one deviating outside the tongue (dashed lines) is always unstable. These two branches coalesce for some detuning value, see Fig.~\ref{ftp}(e). The Turing combs emerging from the intra-branch PDC conditions 
for $\nu=4,5$ are shown in the first two columns in Fig.~\ref{ftp}, 
and the $\nu=16$ cross-branch case is shown in the third column. 

The $\nu=5$ tongue in Fig.~\ref{ftp}(b) corresponds to the  $\om_p+\om_{-5\f}-\om_{-5\s}=0$ sum-frequency matching, see Eq.~\bref{vd}, and therefore, the power of the $-5$th sideband  in the 2nd harmonic is either comparable or even 
exceeds the one in the fundamental, see Figs.~\ref{ftp}(e), (h), and compare the y-axis scales in (d),(e), and (f). The bottom line in Fig.~\ref{ftp} shows how the Turing patterns loop around the resonator in the reference frame rotating with $D_{1\f}$, so that the most efficient generation of the 2nd harmonic leads to the largest differences between the linear and nonlinear  repetition rates, $D_{1\f}-D_{1\nu}$, cf., the pattern angles in Figs.~\ref{ftp}(j), (k), and (l).

Figure~\ref{tur1} shows a more global outlook on the frequency conversion processes happening 
across the PDC tongues when the laser frequency is scanned and its power is fixed, and $\ep_0>0$. 
The stable Turing-combs are typically generated for the relatively small intra-resonator powers achieved for  large positive detunings. When detunings are reduced and the intra-resonator powers are increasing the instabilities bring the breather states producing the denser combs as in Figs.~\ref{tur1}(f),(g),(i), and then  more developed chaotic states as in Figs.~\ref{tur1}(d), (e). 

The difference between the breathers and chaos is further elucidated in Fig.~\ref{tur2}. 
The first and 2nd columns in Fig.~\ref{tur2} compare the spatio-temporal and spatial profiles of the two. 
The coherence of the breather and the incoherence of the turbulent state are confirmed by comparing the RF spectra of the net powers (3rd column in Fig.~\ref{tur2}), 
and the per-mode RF spectra (4th column). The latter shows $|FFT\{\psi_{\mu\f}(t)\}|^2$ plotted vs the mode number, $\mu$, and the RF frequency. The top, i.e., the breather, panel has five visible tilted lines that are vertically separated by the breather period. These are the five sub-combs having different offset frequencies. 

In the example shown in the top line of Fig.~\ref{tur2}, the maximal parametric gain comes to the sidebands
with $\mu=\pm 15$ and to the associated sub-combs, see the top panel in the last column, so that the breather could be interpreted as a quasi-soliton created via the non-degenerate PDC and dominated by the three groups of modes centred around $\mu=-15,0,15$.
The power and per-mode RF spectra in the 2nd line of Fig.~\ref{tur2} are unambiguous about the absence of the inter-mode coherence in the turbulent state.
For studies of the transitions between  the breather and chaotic states associated with the rogue-wave turbulence in the Kerr resonators, see, e.g., Refs.~\cite{matsko1,prxkud,prache}. 

\begin{figure}[t]
	\centering
	\includegraphics[width=0.49\textwidth]{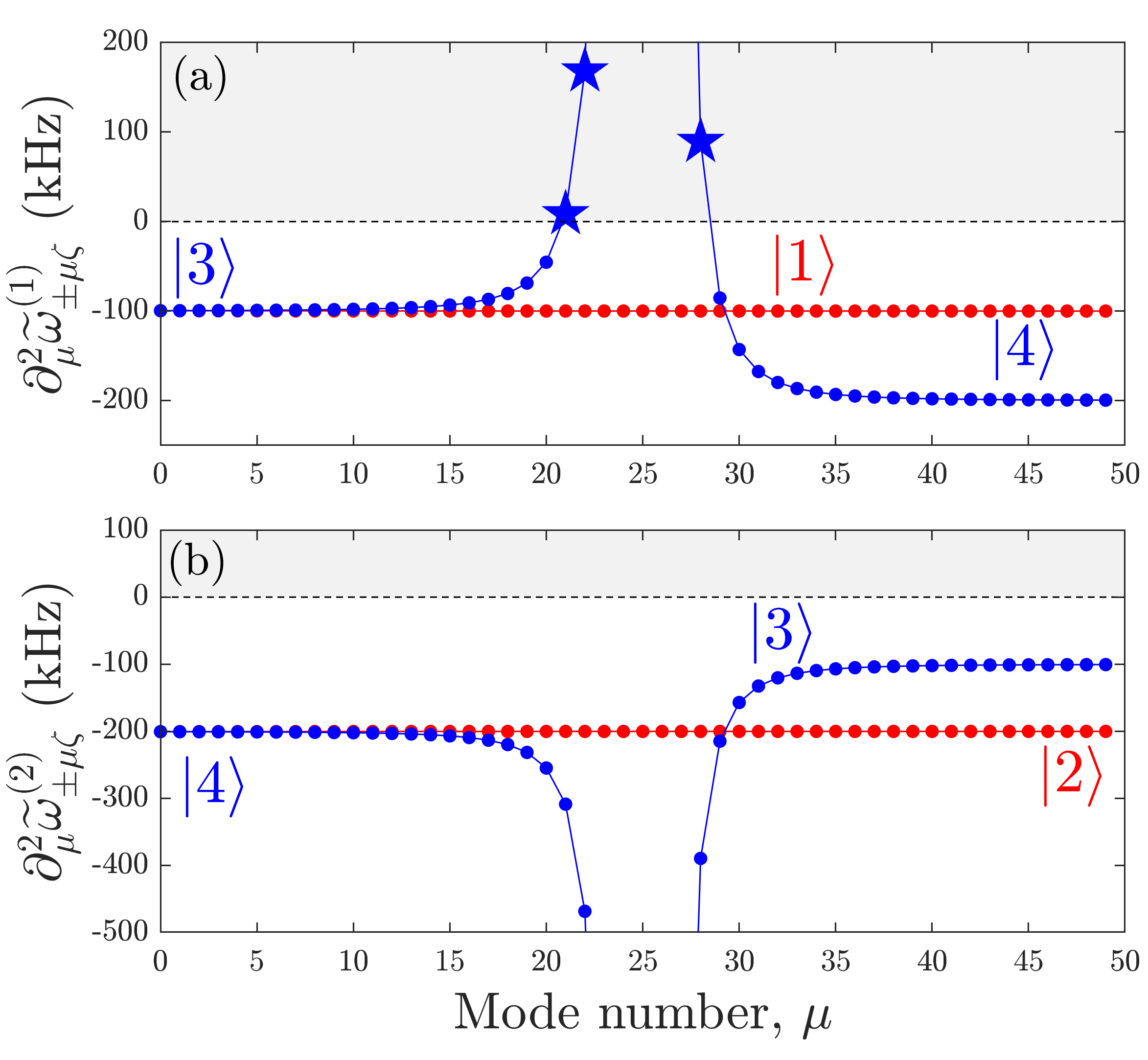}
	\caption{Dispersion of the dressed states, see Eq.~\bref{dsp1}, normalised to $2\pi$:
		(a)  $\wt\om^{(1)}_{\pm\mu\zeta}$, and (b) $\wt\om^{(2)}_{\pm\mu\zeta}$. 
		The red lines correspond to the positive sidebands, $\wt\om^{(1),(2)}_{\mu\zeta}$, and the blue ones to the negative sidebands, $\wt\om^{(1),(2)}_{-\mu\zeta}$.   $-100$kHz, and $-200$kHz are the values of $D_{2\f}$ and $D_{2\s}$  in the bare resonator.  The grey shading and stars show the dressed states with the anomalous dispersion induced by the dressing. In (a), the maximal anomalous dispersion of 
		$\sim 1$GHz (not shown) is achieved at $\mu=25$ for the $\ket{b_{25}^{(3)}}\approx\ket{3}e^{i\phi}-\ket{4}$ state. 
		Parameters are $|\Omega|/2\pi=116$MHz, $\ep_0/2\pi=25$GHz, $\delta=-3.8\kappa_\f$.
	}
	\lab{fnon}
\end{figure}

\begin{figure}[t]
	\centering
	\includegraphics[width=0.49\textwidth]{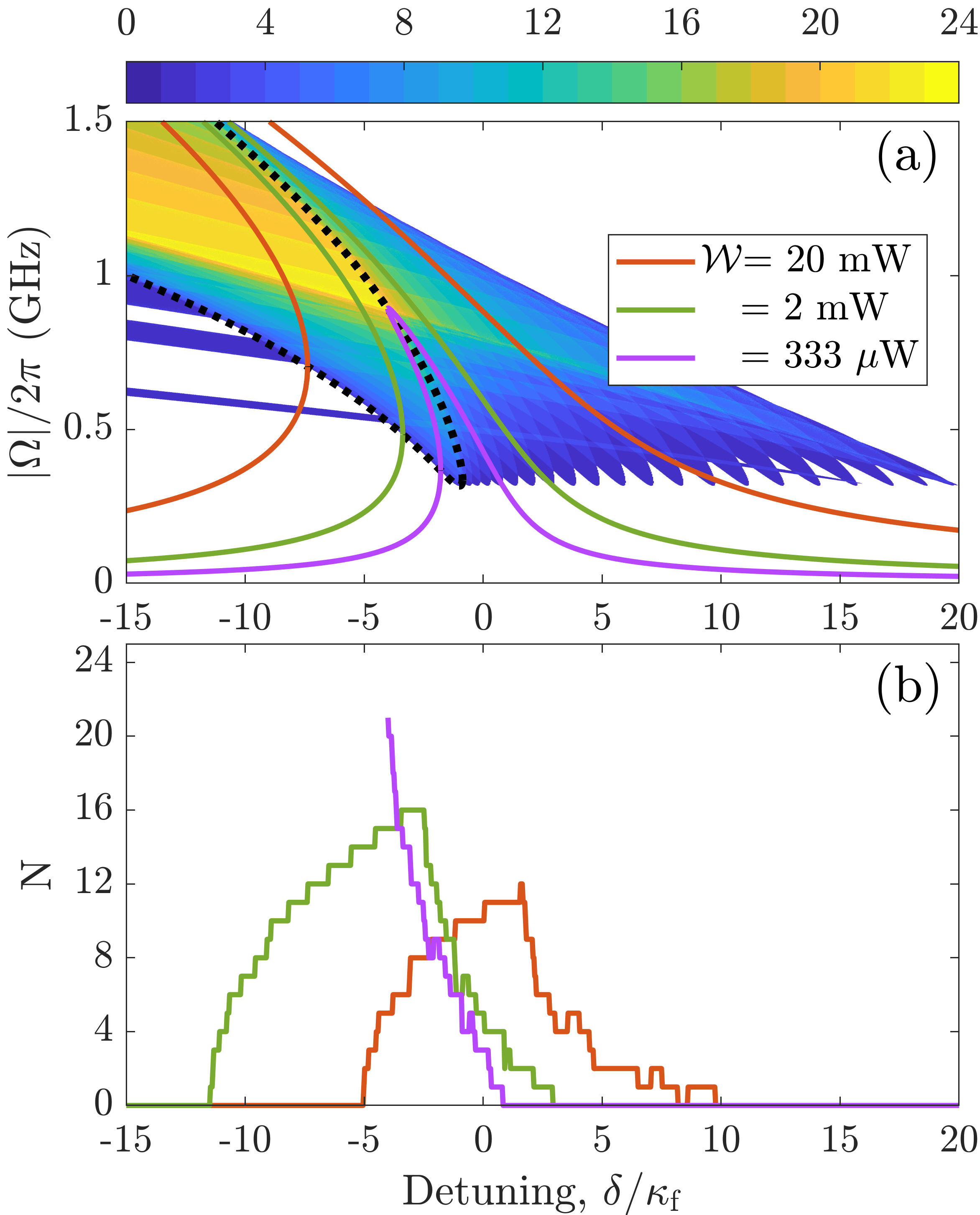}
	\caption{(a) 
		Parametric instability tongues in the ($\delta$,$|\Omega|$) plane, for  $\ep_0/2\pi=25$GHz, cf., $\ep_0/2\pi=5$GHz in Fig.~\ref{ftong}(a). The colorbar shows the number of the simultaneously unstable sidebands, $N$. The black doted line embraces the $\mu=0$ instability range, i.e., the middle branch of the bistability loop. $N$ is clamped by $\mu_*=|\ep_0|/|D_{1\f}-D_{1\s}|=25$, $N\le \mu_*$. (b)  $N$ vs $\delta$ for three values of $\cW$ (laser power) computed along the upper branch of the cw-state.  The magenta line ($\cW=333\mu$W) terminates at the tip of the cw resonance, i.e., at the end of the bistability range. The other two lines are for the higher powers and terminate before the tip, thereby marking stabilisation of the cw-state.
	}
	\lab{sol1}
\end{figure}
\begin{figure*}[t]
	\centering
	\includegraphics[width=1.\textwidth]{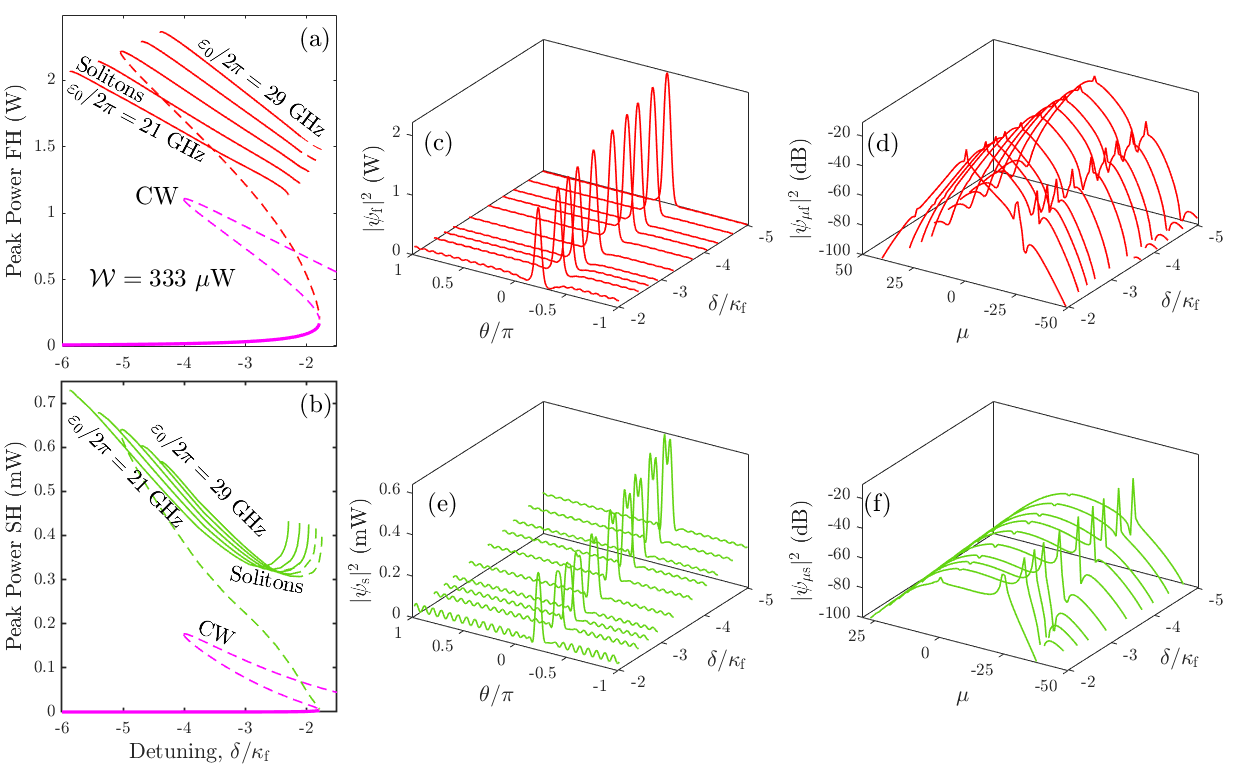}
	\caption{Families of the bright solitons computed for the laser power 
		$\cW=333\mu$W and plotted vs $\delta$. The first line data show the fundamental field,
		and the second line is for the second harmonic. Panels (a,b) show  the soliton branches (red and green lines) and the cw-state (magenta lines) vs $\delta$.  The solitons are shown for  a range of frequency mismatch parameters, $\ep_0/2\pi=$ 21, 23, 25, 27, and 29GHz. 
		The full lines correspond to the stable solutions, and the  dashed lines to the unstable ones.
		The $\ep_0/2\pi=25$GHz case in (a) and (b) is the one that shows how the unstable soliton splits from the cw-state and then connects to the stable soliton.  (c,e) shows how the spatial profile of the soliton changes with $\delta$ for $\ep_0/2\pi=25$GHz. (d,f) are like (c,e) but show the envelopes of the discrete soliton spectra, cf., Fig.~\ref{sol3}. 
	}
	\lab{sol2}
\end{figure*}

\begin{figure}[t]
	\centering
	\includegraphics[width=0.48\textwidth]{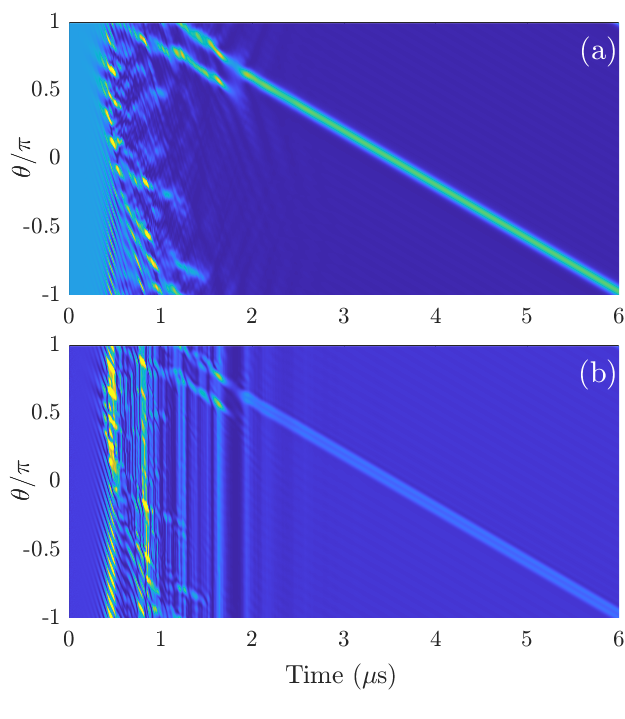}
	\caption{Instability of the cw-state and spontaneous birth of the bright soliton. Laser power $\cW=333\mu$W,  $\ep_0/2\pi=25$GHz, and   $\delta=-3.8\kappa_\f$.
		(a) is the fundamental field and (b) is the second harmonic.
	}
	\lab{sol4}
\end{figure}

\begin{figure*}[t]
	\centering
	\includegraphics[width=1.\textwidth]{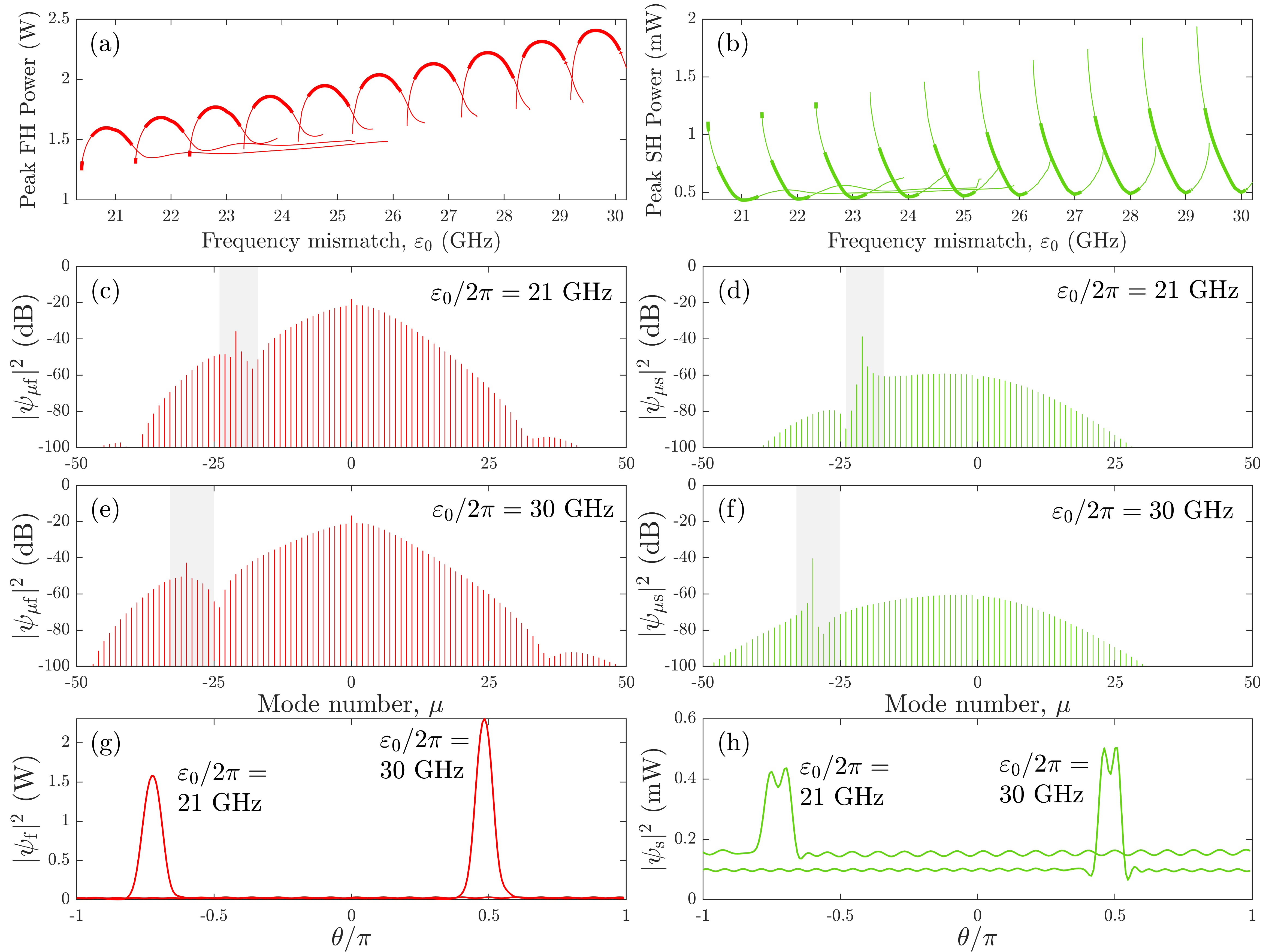}
	\caption{Families of the bright solitons computed for the laser power 
		$\cW=333\mu$W and plotted vs $\ep_0$ for  $\delta=-3.9\kappa_\f$. The first and second columns show the  fundamental and second harmonic data, respectively. Panels (a,b) show the multiple soliton branches centred around  $\ep_0/2\pi\approx \mu_*$, see Eq.~\bref{epe}. (c,d,e,f) show the spectra at the indicated  values of $\ep_0$. (g,i) show the respective spatial soliton profiles. The shaded areas in (c,d,e,f) highlight the  interval between the two zero dispersion points, cf., Fig.~\ref{fnon}. 
	}
	\lab{sol3}
\end{figure*}

\section{Bright soliton prerequisites}
\lab{sec11}

The strong-coupling regime and the dressed state that naturally emerge from it, imply that the parametric gain is relatively small, and therefore, to the leading order, the soliton existence problem 
could be approached through the prism of the nonlinearity vs dispersion balance. 
In this section, we work through the dispersive and nonlinear properties of the dressed states, and the stability of the cw state that underpins the existence of the bright soliton frequency combs. For normal dispersion, $D_{2\zeta}<0$, as we have here, these properties suggest the soliton existence for $\ep_0>0$, which is the case we describe in details below. Solitons themselves are introduced in Section~\ref{sec12}.

\subsection{Dispersion of dressed states}

The microresonator dressed states are parametrised by momentum $\mu$, and represent families  of the quasi-particles, photon-photon polaritons, with the effective mass inversely proportional to the polariton dispersion~\cite{prr}. Our focus here is on the frequency conversion, and  therefore the dispersion terminology is more natural. 
Dispersions of the first and second branches in the dressed spectrum  are calculated as
\bsub
\lab{dsp1}
\begin{align}
&	\p^2_\mu\wt\om_{\pm\mu\zeta}^{(1)}=\p^2_\mu\beta_{\pm\mu}^{(1)}
	=\tfrac{1}{2}(D_{2\f}+D_{2\s})+\tfrac{1}{2}\Omega''_{\pm\mu},
\lab{dsp1a}\\
&	\p^2_\mu\wt\om_{\pm\mu\zeta}^{(2)}=\p^2_\mu\beta_{\pm\mu}^{(2)}
=\tfrac{1}{2}(D_{2\f}+D_{2\s})-\tfrac{1}{2}\Omega''_{\pm\mu},
\lab{dsp1b}
\end{align}
\esub
where, $\Omega''_{\mu}=\Omega_{\mu+1}
	+\Omega_{\mu-1}-2\Omega_{\mu}$.  $\Omega''_{\mu}$ depends on $|\Omega|$ (i.e., on the sum-frequency nonlinearity),  $D_{2\zeta}$, and on the repetition rate difference, $D_{1\f}-D_{1\s}$, which means that the dressed state dispersion is determined by the interplay of all three factors. This is unlike the  bare-state dispersion, trivially given by $\p^2_\mu\om_{\pm\mu\zeta}=D_{2\zeta}$.

For $\ep_0>0$, none of the $\wt\om_{\mu\zeta}^{(1)}$ frequencies is sum-frequency matched, while $\wt\om_{-\mu\zeta}^{(1)}$ are quasi-matched for $\mu$ around $\mu_*$, see Section~\ref{sec4} and Fig.~\ref{frabi0}. Therefore, the dressed states corresponding to $\wt\om_{\mu\zeta}^{(1)}$ are the quasi-bare states, $\ket{b_\mu^{(1)}}\approx\ket{1}$, with the dispersion $\approx D_{2\f}$, see the red line in Fig.~\ref{fnon}(a).
The dressed states corresponding to $\wt\om_{-\mu\zeta}^{(1)}$ are  
$\ket{b_{-\mu}^{(1)}}\approx\ket{3}$ for $\mu<\mu_*$, then, in the proximity of $\mu_*$, they hybridise to the maximally dressed state,
$\ket{3}e^{i\phi}-\ket{4}$, and then transform to the quasi-bare $\ket{4}$ states.

It is instructive to evaluate the dressed dispersion for $\mu$ around $\mu_*$, where
$\vD_{-\mu}\approx 0$, $\Omega_{-\mu}\approx|\Omega|$, see Eqs.~\bref{ra00}, \bref{del}, 
\be
\begin{split}
	\Omega''_{-\mu}
	&\approx\frac{\vD_{-\mu}^3\vD_{-\mu}''+|\Omega|^2\big(\vD'_{-\mu}\vD_{-\mu}\big)'}{\Omega_{-\mu}^3} \\
	&\approx  \frac{(\vD'_{-\mu})^2}{|\Omega|}
	\approx
	\frac{(D_{1\f}-D_{1\s})^2}{|\Omega|}\sim 1\text{~to~}10\text{GHz}.
\end{split}
\lab{dsp2}
\ee
The above strongly dominates over $D_{2\zeta}$, so that,
the dispersion of $\wt\om_{-\mu\zeta}^{(1)}$ becomes large anomalous. Thus, dressing the $\wt\om_{-\mu\zeta}^{(1)}$ branch  
creates two zero dispersion points around $\mu_*$ and puts a group of modes into the range of anomalous dispersion, see Fig.~\ref{fnon}(a). 

The dressing induced dispersion changes are very large, see Eq.~\bref{dsp2},
so that for the nonlinearity to compensate the dispersive pulse 
spreading  it would be preferential to engage the modes with the 
bare resonator dispersion around $\mu=0$,
and away from $\mu_*$. Thus keeping $\mu_*\gg 1$ and increasing it by tuning the index 
to make  $|\ep_0|$ larger, see Eq.~\bref{epe}, is expected to 
increase the bandwidth of modes suitable for the bright soliton modelocking regime.

In other words, the  repetition rate difference can be large, but the sum-frequency matching to the 
appropriately large sideband number puts the undesirably large dispersion 
this difference creates away from the spectral core of the soliton.

\subsection{Optical Pockels and cascaded-Kerr nonlinearities}
Since we are expecting to work with the normal dispersion, the bright solitons 
would require the negative, i.e., defocusing, nonlinearity to compensate for it.
Derivatives of $\wt\om_{\pm\mu\zeta}^{(j)}$ in $|\Omega|$ describe 
the rate of the nonlinear shifts of the frequencies in the dressed 
spectrum with the intra-resonator power,
\be
\begin{split}
	\frac{\p\wt\om_{\pm\mu\zeta}^{(1)}}{\p |\Omega|}&=
	\frac{1}{2}\frac{\p\Omega_{\pm\mu}}{\p |\Omega|}=\frac{|\Omega|}{2\Omega_{\pm\mu}}>0,\\
	\frac{\p\wt\om_{\pm\mu\zeta}^{(2)}}{\p |\Omega|}&=-
	\frac{|\Omega|}{2\Omega_{\pm\mu}}<0.
\end{split}
\lab{shif}
\ee
The signs of the above expressions determine the signs of 
the effective nonlinearities experienced by the sidebands.
Since the resonator frequencies are inversely proportional to the refractive index, see Eq.~\bref{refin}, we conclude that the frequencies $\wt\om_{\pm\mu\zeta}^{(2)}$
in the dressed spectrum experience the net positive (self-focusing) nonlinearity, and
$\wt\om_{\pm\mu\zeta}^{(1)}$ - the net negative (defocusing) nonlinearity.

Though the branch nonlinearities do not change  signs, their dependence on $|\Omega|$ varies profoundly. Indeed, $\Omega_{\pm\mu}$ admits two qualitatively different Taylor series expansions,
see Table~\ref{t2},
\bsub
\lab{r3}
\begin{align}
	&\Omega_{\pm\mu}=|\Omega|+\frac{\vD_{\pm\mu}^2}{2|\Omega|}+\dots,
	~\frac{\vD_{\pm\mu}^2}{|\Omega|^2}\ll 1,
	\lab{r3a}\\
	&\Omega_{\pm\mu}=|\vD_{\pm\mu}|+\frac{|\Omega|^2}{2|\vD_{\pm\mu}|}+\dots,~
	\frac{\vD_{\pm\mu}^2}{|\Omega|^2}\gg 1.
	\lab{r3b}
\end{align}
\esub
Thus, if $\mu$ is near to $\mu_*$, then 
the nonlinear shift of $\wt\om_{\pm\mu\zeta}^{(j)}$ frequency is proportional to the amplitude of the cw-state, $|\Omega|\sim |\psi_\f|$, which corresponds to the optical Pockels effect, see Eq.~\bref{r3a}.  While, for $\mu$ away from $\mu_*$, the  nonlinear shifts are proportional to the power, $|\Omega|^2\sim |\psi_\f|^2$, and hence are the Kerr-like, i.e., correspond to the cascaded-Kerr effect, see Eq.~\bref{r3b}.

\subsection{CW-state stability and instability}
From the above, one should conclude that, for $\ep_0>0$, the first branch of the dressed 
states should be considered as the one able to form the bright solitons. Since it provides  a combination of the defocusing nonlinearity and  the relatively small normal dispersion, 
except several $\mu$'s around $\mu_*$, where  dispersion is anomalous.

The same condition, as the one just stated, leads to the intra-branch instability of the cw-state,
see and compare Figs.~\ref{ftong}(a) for $\ep_0/2\pi=5$GHz, and  \ref{sol1}(a) for $\ep_0/2\pi=25$GHz. 
Recalling the results of Section~\ref{sec6}, see Eq.~\bref{e47}, the number, $N$, of the sidebands that can be simultaneously unstable under these conditions  is clamped by $\mu_*$. Therefore, and also in line with the previous subsection, $\ep_0/2\pi=25$GHz brings more of the unstable sidebands.  
Fig.~\ref{sol1}  shows that there exists the optimal laser power, $\cW$, achieving the maximal $N$. 

$N$ going up and then down with $\delta$ tuned more negative and  powers increasing, see Fig.~\ref{sol1}(b), is due to the nonlinear shifts becoming saturated by the higher order terms in the expansions of the root function in $\Omega_{\pm\mu}$. The weakened nonlinear shifts gradually bring the parametric gain below the threshold first for some and finally  for all the sidebands.
The $N$ vs $\delta$ dependencies in Fig.~\ref{sol1}(b) would be very different in the Kerr resonators, where the number of the  unstable sidebands tends to infinity with the simultaneous increases of $\delta$ and $\cW$~\cite{pra}.

It is now important to note that, for $\delta<0$, the above discussed instabilities happen to  the upper state of the cw bistability loop, while the low branch is either exclusively or largely stable,  see Fig.~\ref{sol1}(a). The stable low-amplitude cw-state makes the background 
for the bright solitons reported in the next section.

\section{Bright soliton frequency combs}
\lab{sec12}

The results of the previous Section let us to conclude that, for the normal dispersion of the bare resonator modes, $D_{2\zeta}<0$,
the bright solitons are expected providing one arranges the 
index/frequency matching parameter between the $M$ and $2M$ resonator modes to be positive,
\be
\ep_0\sim (n_{2M}-n_M)>0,
\ee
and sufficiently large, so that the 
sum-frequency matching or near-matching, 
\be
\om_{0\f}+\om_{-\mu\f}-\om_{-\mu\s}=0,
\ee
happens for
\be
\mu=\mu_*\gg 1.
\lab{solcon}
\ee 
The detuning should than be tuned to $\delta<0$, i.e., 'blue' detuning. 
$\mu_*$ is well approximated by the ratio between $|\ep_0|$, and the repetition rate difference, $|D_{1\f}-D_{1\s}|$, see Eq.~\bref{epe}.

We now do as prescribed and select the laser power, $\cW$, and detuning, $\delta$, corresponding to a large number of the simultaneously unstable 
sidebands, see Fig.~\ref{sol1}. Initialising Eq.~\eqref{mm} with the cw-state we immediately observe its instability and the subsequent formation of the stable two-colour soliton, see Fig.~\ref{sol4}.

For $D_{2\zeta}>0$ (anomalous dispersion),
the bright solitons would require $\ep_0<0$, 
adjusting $\om_{0\f}+\om_{\mu\f}-\om_{\mu\s}=0$ for $\mu=\mu_*\gg 1$, and, then, tuning to 
$\delta>0$, i.e., 'red' detuning. The mixed dispersion case, $D_{2\f}D_{2\s}<0$, requires a separate analysis.

To trace the soliton families in the parameter space we solve the comb equations in Section~\ref{sec9}. Since the teeth of the single soliton combs follow with step one, i.e., their spatial period is $2\pi$, we set $\nu=1$ in Eqs.~\eqref{tp}, \eqref{tp1}. 
Families of the bright solitons traced in $\delta$ for the laser power 
$\cW=333\mu$W and  for a range of frequency mismatch parameters, $\ep_0/2\pi\in[21,29]$GHz are shown in Fig.~\ref{sol2}. 
The unstable soliton branch (dashed red and green lines) split from the middle branch of cw state (dashed magenta lines) at the point of its $\mu=1$ instability. The stable soliton branches extend outside the cw-bistability towards more negative $\delta$'s.

The soliton profiles along the resonator circumference shown in Figs.~\ref{sol2}, \ref{sol3} are characterised by the tails oscillating with the period $2\pi/\mu_*$. The corresponding spectra have pronounced peaks at $\mu=-\mu_*$, where the powers of the fundamental and 2nd harmonic sidebands are balanced due to the sum-frequency matching. 
The background of the soliton in the fundamental field is set primarily by the $\mu=0$ sideband, 
see spectra in Fig.~\ref{sol2}(d), where $\mu=0$ is stronger than $\mu=-\mu_*$. Therefore, the fundamental background drops with $\delta$ becoming more negative, cf., the correlated changes of the full CW line in Fig.~\ref{sol2}(a) and of the soliton background in Fig.~\ref{sol2}(c). The background of the soliton in the 2nd harmonic is, however, set primarily by the $\mu=-\mu_*$ sideband, see spectra in Fig.~\ref{sol2}(f), where $\mu=-\mu_*$ is the strongest and its power correlates with the soliton peak power. This explains why the 2nd harmonic soliton background goes up with $\delta$ becoming more negative, cf., the dropping full CW line in Fig.~\ref{sol2}(b) and the increasing soliton background in Fig.~\ref{sol2}(e).

The oscillatory soliton tail should be interpreted as due to the inability of the nonlinear effects to compensate for the sharp rise and the sign change of the dressed state dispersion around $\mu_*$, see Fig.~\ref{fnon}. Therefore, the nature of the tails here is similar to the  soliton Cherenkov radiation 
in the resonators~\cite{pral,mil0,kip11}, fibers~\cite{rmp}, and bulk crystals~\cite{oe1}. 
We note, that, in the present case, the dispersion can be significantly altered by the pump power dependent state dressing, which provides a more flexible tool to control the  radiation frequency.

Figure~\ref{sol3} shows how the soliton families change with the tuning of  $\ep_0$. Here the structure of the families appears to be more complex. The period of the oscillations of the soliton tail, $2\pi/\mu_*$, can change only discretely  while  $|\ep_0|$ is tuned continuously, so that the solution is forced to accommodate itself, as much as it can, to the rigid period of its tails, which leads to the discrete set of the soliton families. Each family is centred around the value of $|\ep_0|=\mu_*|D_{1\f}-D_{1\s}|$, we recall that $|D_{1\f}-D_{1\s}|/2\pi=1$GHz. The tail oscillations of the Kerr solitons in microresonators with the large higher-order dispersions also lead to somewhat similar 'quantized' behaviour of the soliton parameters~\cite{mil}.

\section{Discussion}
There are numerous open problems left for a researcher tempted to understand the multimode dynamics of the high-Q $\chi^{(2)}$ microresonators by looking into properties of the individual modes, in line with the present-day experimental capabilities. 
The extension of our results to the case of the exact index matching 
$\ep_0=0$ while keeping  $D_{1\f}-D_{1\s}$  large requires separate consideration. 
Extending the bandwidth of the soliton combs by taking the shorter resonators with the higher repetition rates and lower quality factors, and, perhaps, larger $\mu_*$ needs to be investigated. 

The power induced dispersion engineering of the dressed states offers a new method to control  the comb widths and shapes. Observations of the solitons, the Rabi splitting and the sum-frequency matching associated spectral features predicted here are of course important and can be attempted with the available resonators and the index matching control tools. We note, the relatively high conversion efficiency from the pump to the soliton-comb spectra seen in Fig.~\ref{sol3}, which could be an important practical aspect requiring further exploration.

We should recall here the  prior theoretical work
on the spatial~\cite{ol2,etr} and temporal~\cite{wabol,av} 
resonator solitons achieved via the 2nd harmonic generation arrangements requiring 
the exactly or near matched phase, $\ep_0=0$, and group, $D_{1\f}=D_{1\s}$, velocities. 
While a combination of these assumptions appears as the desirable idealisation in the contexts 
of the currently available ring microresonators, future studies along these lines are warranted.
The relevance of the strong-coupling and dressed states for the microresonator 
half-harmonic generation arrangement remains to be analysed, including their
links to the results on the half-harmonic 
bright-bright, dark and dark-bright soliton pulses~\cite{bru,ol,preold,skrd,sim,lob,pra2,wlo}.

\section{Summary}

(1) The theoretical framework, i.e., dressed-resonator method, to study frequency conversion and solitons  is formulated by including the sum-frequency nonlinearity into the definition of the resonator spectrum. 

(2) The Rabi splitting of the dressed frequencies leads to the four distinct PDC conditions, see Eq.~\bref{z0},
which are used to explain the existence and generation of the sparse non-soliton, i.e., Turing-pattern-like, frequency combs.

(3) The effective nonlinearity and dispersion of the dressed states have been used to demonstrate 
that the microresonator with the normal dispersion and 
naturally  large difference of the repetition rates at the fundamental and 2nd harmonic frequencies, $D_{1\f}-D_{1\s}$, supports a family of the bright soliton frequency combs, see Figs.~\ref{sol2}, \ref{sol3}. Conditions for this are provided  by tuning  the index/frequency matching parameter, $\ep_0=2\om_{0\f}-\om_{0\s}$, to be positive and large, so that it exceeds the repetition rate difference by a significant factor, $\mu_*=|\ep_0|/|D_{1\f}-D_{1\s}|\gg 1$. $\mu_*$ or $-\mu_*$ approximate the mode number associated with the phase-matched sum-frequency process and set limits on the soliton bandwidth.

\section{Acknowledgement}
This work was supported by the EU Horizon 2020 Framework Programme (812818, MICROCOMB).

\appendix
\section*{Appendices}
\section{Envelope equations}\lab{ap1}
The intra-resonator electric fields of the fundamental and 2nd harmonic are expressed as
per Eq.~\bref{field}.
Envelopes of the fundamental, $\psi_\f$, and second, $\psi_\s$, harmonic  
satisfy the following system of the partial-differential equations
\bsub
\lab{mm}
\begin{align}
	i\p_t \psi_\f &=\delta \psi_\f-iD_{1\f}\p_\vta \psi_\f-\tfrac{1}{2}D_{2\f}\p^2_\vta \psi_\f \lab{mma}\\
	&-\gamma_{\f} \psi_\s \psi_\f^* -\cN_{\f}-i\tfrac{1}{2}\kappa_\f\big(\psi_\f-\cH\big)\nn
	,\\ 
	i\p_t \psi_\s&=(2\delta-\ep_0) \psi_s-iD_{1\s}\p_\vta \psi_\s-
	\tfrac{1}{2}D_{2\s}\p^2_\vta \psi_\s \lab{x}\\
	&-\gamma_{\s} \psi_\f^2 -\cN_{\s}-i\tfrac{1}{2}\kappa_\s \psi_\s.\nn
\end{align} 
\esub
Conditions $\psi_{\zeta}(t,\vta)=\psi_{\zeta}(t,\vta+2\pi)$
make this system equivalent to a set of the coupled-mode equations for $\psi_{\mu\zeta}(t)$. The first principle derivation of Eq.~\bref{mm} is given in Ref.~\cite{josab}.

All parameters are explained in Section~\ref{sec2} of the main text, apart from  
$\cH^2$, which characterizes the  pump power~\cite{josab}.
If $\cF=D_{1\f}/\kappa_\f=20000$  is the finesse, then $\cH^2$ is expressed via the 
incoming laser power $\cW$ as 
\be
\cH^2=\frac{\eta}{\pi} \cF\cW.
\lab{pow}
\ee
$\eta<1$ is the coupling efficiency.
$\cN_{\zeta}$ are the intrinsic Kerr, i.e. $\chi^{(3)}$, nonlinearity terms, 
\be
\cN_{\f,\s}=\gamma_{3\f,3\s}\big(|\psi_{\f,\s}|^2+2|\psi_{\s,\f}|^2\big)
\psi_{\f,\s}.
\lab{ke}
\ee
Nonlinear coefficients $\gamma_{\zeta}/2\pi$ and $\gamma_{3\zeta}/2\pi$ have units 
of Hz$/\sq{\text{W}}$ and $\text{Hz}/{\text{W}}$, respectively~\cite{josab}.
Units and numerical values of other parameters can be found in Table~\ref{t1}.

\section{CW-state: $\chi^{(2)}$ vs $\chi^{(3)}$}\lab{ap2}
CW state, i.e.,  the $\mu=0$ mode in the fundamental
and its 2nd harmonic, is a solution  of Eqs.~\bref{mm} 
with $\psi_{\zeta}= \psi_{0\zeta}$, $\p_\vta\psi_{0\zeta}=\p_t\psi_{0\zeta}=0$.
Let us now evaluate the relative impact of the $\chi^{(2)}$ and $\chi^{(3)}$ effects on the cw state. 
If $\ep_0$ dominates over the linewidth and detuning parameters, then 
$\psi_{0\s}\approx -\gamma_{\s}\psi_{0\f}^2/\ep_0$, and
the net nonlinear frequency shift of the  fundamental resonance is  
\be
\left[-\frac{\gamma_{\s}\gamma_{\f}}{\ep_0}+\gamma_{3\f}\right]|\psi_{0\f}|^2=
\gamma_{3\f}\left[-\frac{\ep_{\text{cr}}}{\ep_0}+1\right]|\psi_{0\f}|^2.
\lab{ck}
\ee
Hence, only for $|\ep_0|\gtrsim |\ep_{\text{cr}}|$, $\ep_{\text{cr}}=\gamma_{\s}\gamma_{\f}/\gamma_{3\f}$,
the 2nd harmonic becomes weak enough for the 
$\chi^{(3)}$ induced shift to catch up with the $\chi^{(2)}$ one.
For the parameters in Table~\ref{t1} and 
$\gamma_{3\zeta}/2\pi\lesssim 1\text{MHz}/{\text{W}}$~\cite{pra},
$\ep_{\text {cr}}/2\pi$ is $\gtrsim 100$GHz.
Thus, in the range of $\ep_0/2\pi\in [-30,30]$GHz  explored in this project 
and for the combs with the relatively low powers,  the $\chi^{(3)}$ terms can be neglected.
However, most of the numerical data in this work  have been calculated with and without  $\cN_{\zeta}$. The differences that we have observed  are not worth mentioning in the context of our study. For all the above reasons,  we set $\cN_{\zeta}=0$. Some results on the interplay of the $\chi^{(2)}$ and $\chi^{(3)}$ effects 
in the half-harmonic generation setup can be found in, e.g., Refs.~\cite{ol,bru}.

\section{CW-state: $\chi^{(2)}$ only}\lab{ap3}
The cw state is sought in the form
\be
\psi_{0\f}=
\frac{\Omega_{~}}{\sqrt{2\kappa_\f\kappa_\s}}\cH_*,~\psi_{0\s}=\frac{8\gamma_{2\s}\psi_{0\f}^2}{\Omega_\s}=\frac{\Omega^2}{\gamma_{2\f}\Omega_\s},
\lab{nd}
\ee
where $\Omega$ is the Rabi frequency.
$\Omega_\s$ is defined in Eq.~\bref{os},
and \be
\cH^2_*=\frac{\kappa_\f\kappa_\s}{4\gamma_{2\f}\gamma_{2\s}}
\approx 30\mu\text{W}
\lab{star0}
\ee
 is the characteristic intra-resonator power. Thus, the explicit relation between the Rabi frequency and the power in the fundamental field is
\be
|\psi_{0\f}|^2=
\frac{|\Omega|^2}{8\gamma_{2\f}\gamma_{2\s}}.
\lab{ws1}
\ee 

Using Eqs.~\bref{mm},  one can show that
\be
\Omega=\sqrt{\frac{\kappa_\f\kappa_\s}{2}}\sqrt{\frac{\cW~}{\cW_*}}
\frac{\kappa_\f}{\Omega_\f}\left[
1-\frac{|\Omega|^2}{\Omega_\f\Omega_\s}\right]^{-1}.\lab{cw0}
\ee
Here, $\Omega_{\f,\s}$ are defined after Eq.~\bref{cw8}, and 
\be
\cW_*=\frac{\pi\cH_*^2}{\eta\cF}\approx 10\text{nW}.
\lab{star} 
\ee
Taking the modulus squared of Eq.~\bref{cw0} we find the real cubic equation for $|\Omega|^2$,
\begin{align}
	\frac{2|\Omega_\f|^2|\Omega|^2}{\kappa_\f^3\kappa_\s}
	\times&\left[1-|\Omega|^2\text{Re}\left\{\frac{2}{\Omega_\f\Omega_\s}\right\}
	+\frac{|\Omega|^4}{|\Omega_\f|^2|\Omega_\s|^2}\right]
	\nn \\
	=&\frac{\cW}{\cW_*}=\frac{\cH^2}{\cH_*^2}.
	\lab{cw00}
\end{align}

A useful insight into the  cw properties is provided by taking the limit when
$|\ep_0|$ is large relative to $|\delta|$, and $|\delta|$ is large relative to $\kappa_\zeta$, so that
$\Omega_\s\approx -8\ep_0$ and  $\Omega_\f\approx \delta$. Then, bistability  of the cw-state
requires $\delta\ep_0<0$ (the square bracket in Eq.~\bref{cw0} can be zero). In this regime, 
\be
\max_\delta|\Omega|^2\simeq -\delta\ep_0,
\lab{bis}
\ee 
 implying that 
the resonance shifts proportionally to the pump power, i.e., in the same way as it happens in Kerr effect.
It means, that the cw-state behaves as it would in the Kerr resonator. 
For $\ep_0=0$ or small, 
$\Omega_{\f}\approx \delta$, $\Omega_{\s}\approx 16\delta$. Then, the  cw becomes simultaneously bistable for 
$\delta>0$ and $\delta<0$ \cite{wabol,av}, and $\max_\delta|\Omega|^2\simeq\delta^2$, 
i.e., 
\be
\max_\delta|\Omega|\simeq |\delta|.
\ee
This is the Pockels regime of the cw-state, when the nonlinear change of the refractive index is proportional to the first power of the field amplitude, which we do not consider in this work. Conditions for the $\mu\ne 0$ sidebands to be in either cascaded-Kerr or Pockels regimes are different and discussed in Section~\ref{sec11}~B.

\section{Linearization around the cw-state}\lab{ap4}
In order to develop a theory of the growth of the  multi-sideband signal, i.e., frequency comb,  on top of the cw  solution, $\psi_{0\zeta}$, we extend Eqs.~\bref{mm} by a pair of the complex-conjugated equations~\cite{skrold}, and seek a solution in the form
\be
\begin{split}
&\begin{bmatrix}
		\psi_{\f}\\
		\psi_{\s}\\
		\psi^*_{\f}\\
		\psi^*_{\s}
\end{bmatrix}=\begin{bmatrix}
\psi_{0\f}\\
	\psi_{0\s}\\
	\psi^*_{0\f}\\
	\psi^*_{0\s}
\end{bmatrix}+	\\
&\sum_{\mu\ge 0}\left\{\begin{bmatrix}
\wt	\psi_{\mu\f}\\
\wt	\psi_{\mu\s}\\
\wt	\psi_{-\mu\f}\\
\wt	\psi_{-\mu\s}
\end{bmatrix}e^{i\mu\ta}
+\begin{bmatrix}
	0 & 0 & 1 & 0\\
	0 & 0 & 0 & 1\\
	1 & 0 & 0 & 0\\
	0 & 1 & 0 & 0
\end{bmatrix}
\begin{bmatrix}
\wt	\psi_{\mu\f}^*\\
\wt	\psi_{\mu\s}^*\\
\wt	\psi_{-\mu\f}^*\\
\wt	\psi_{-\mu\s}^*
\end{bmatrix}e^{-i\mu\ta}\right\}.
\end{split}
\lab{sw}
\ee
If $\vta$ is the angle measured along the resonator circumference, then $\ta=\vta-D_{1}t$
is the coordinate in the rotating frame, see Eq.~\bref{gal}.
$\wt\psi_{\pm\mu\zeta}(t)$ are the sideband amplitudes.
Summing up the first and second lines gives the net signals in the fundamental and 2nd harmonic, respectively, with the third and fourth lines being their conjugated, see Eq.~\bref{anz}. Substituting Eq.~\bref{sw} into the extended Eq.~\bref{mm}, we assume smallness of the sideband amplitudes, linearise, and then derive equations for the individual sidebands using the angular momentum matching. For a given $\mu$ this leads to a coupled system of the four ordinary differential equations for $\wt\psi_{\mu\f}$,
$\wt\psi_{-\mu\f}$, $\wt\psi_{\mu\s}$, and $\wt\psi_{-\mu\s}$, see Eq.~\bref{kk}.

\section{Approximate PDC conditions}\lab{ap5}
To present the  PDC condition in Eq.~\bref{syn1} in a more transparent form,
we first make explicit how $\delta$, $\ep_0$, and $\mu_*$ are implicated there,
\begin{align}
	|\Omega_{\text{pdc}}|^2&=\frac{4
	}
	{[(\delta-\delta_{\mu\f})+(2\delta-\delta_{\mu\s}-\ep_0)]^2}
	\nn\\
	&\times
	\big(\delta-\delta_{\mu\f}\big) \nn\\
	&\times\big(2\delta-\delta_{\mu\s}-\ep_0\big)
	\nn\\
	&\times
	\big(3\delta-\delta_{\mu\s}-\delta_{\mu\f}
	-\ep_0(1+\dfrac{\mu}{\mu_*})\big)
	\nn\\
	&\times
	\big(3\delta-\delta_{\mu\s}-\delta_{\mu\f}-
	\ep_0(1-\dfrac{\mu}{\mu_*})\big),
	\lab{syn2}
\end{align}
where we have defined
\be
\delta_{\mu\f}=-\tfrac{1}{2}D_{2\f}\mu^2,~\delta_{\mu\s}=-\tfrac{1}{2}D_{2\s}\mu^2.
\ee

Eq.~\bref{in1} implies that $|\ep_0|$ dominates over all $\delta$'s. 
The first case to consider is when $\mu_*$ falls between the two nearest integers, i.e.,
the sum-frequency matching point, $\mu=\mu_*$, has been missed. 
Then Eq.~\bref{syn2} simplifies to Eq.~\bref{kerr0},
\be
|\Omega_{\text{pdc}}|^2\approx -4\ep_0\left(\delta-\delta_{\mu\f}\right)
\left[1-\frac{\mu^2}{\mu_*^2}\right].
\lab{kr0}
\ee 

If,  the sum-frequency process is either nearly or exactly matched at
\be
\mu=\wh\mu\approx\mu_*,~\text{where}~ \wh\mu\in\mathbb{Z},
\mu_*\in\mathbb{R},
\ee 
then 
\be
|\Omega_{\text{pdc}}^{(\wh\mu)}|^2\approx 
8(\delta-\delta_{\wh\mu\f})
\big(3\delta-\delta_{\wh\mu\s}-\delta_{\wh\mu\f}\big).
\lab{po}
\ee
Transition from Eq.~\bref{kr0} to Eq.~\bref{po} implies transition from the square-root (Kerr-like) dependence of $|\Omega_{\text{pdc}}|$ vs $\delta$ to the quasi-linear (Pockels-like) one.

\section{Laser power at the PDC thresholds}\lab{ap6}
Detunings at the tips of the instability tongues are worked out by imposing conditions
\bsub
\begin{align}
&	|\Omega_{\text{pdc}}|^2=|\Omega_{\text{th}}^{(\mu)}|^2,
\lab{f1a}\\
&	|\Omega_{\text{pdc}}^{(\wh\mu)}|^2=|\Omega_{\text{th}}^{(\wh\mu)}|^2.
\lab{f1b}
	\end{align}
\esub

Eqs.~\bref{f1a}, \bref{kerr0}, \bref{th1} yield
\be
\delta_{\text{th}}^{(\mu)}=\delta_{\mu\f}-\frac{\kappa_\f~\text{sgn}(\ep_0)}{1-\dfrac{\mu^2}{\mu_*^2}},~
\mu\ne\wh\mu
\lab{f2}
\ee

For $\mu=\wh\mu$, the procedure is the same. In order not to overcomplicate the answer, we impose a plausible condition $(D_{2\f}+D_{2\s})=3D_{2\f}$, leading to 
$\delta_{\wh\mu\s}+\delta_{\wh\mu\f}=3\delta_{\wh\mu\f}$. Then,
Eqs.~\bref{f1b}, \bref{po}, \bref{th2} yield
\be
\delta_{\text{th}}^{(\wh\mu)}=\delta_{\wh\mu\f}\pm\sqrt{\frac{|\ep_0|}{6}\sqrt{\kappa_\f(\kappa_\f+\kappa_\s)}}.
\lab{f3}
\ee

Transparent analytic estimates for the laser powers $\cW$  at the tips of the instability tongues
can be worked out after observing that along the tails of the nonlinear resonances, see Fig.~\ref{fcw}, 
the square bracket in Eq.~\bref{cw0} is $\approx 1$,
\be
\frac{\cW}{\cW_*}\approx\frac{2\delta^2|\Omega|^2}{\kappa_\f^3\kappa_\s},
\lab{f4}
\ee
see Eq.~\bref{star} for $\cW_*$.

The balance of terms in Eq.~\bref{f3} is such that the root term dominates and 
$\delta_{\wh\mu\f}$ can be neglected, which gives  the following estimate for the  power
\be
\frac{\cW_{\text{th}}^{(\wh\mu)}}{\cW_*}\approx
\frac{4|\ep_0|^2(\kappa_\f+\kappa_\s)}{3\kappa_\f^2\kappa_\s},~\mu=\wh\mu.
\lab{f5}
\ee

 In Eq.~\bref{f2}, the two terms are  balanced
for the moderate $\mu$'s leading to a longer equation not included here, but for $\mu^2\gg\mu_*^2$, the second term can be neglected, so that
\be
\frac{\cW_{\text{th}}^{(\mu)}}{\cW_*}\approx
2\mu^4\frac{D_{2\f}^2|\ep_0|}{\kappa_\f^2\kappa_\s},~\mu\gg\mu_*.
\lab{f6}
\ee

Thus, the powers to generate the sum-frequency matched sideband, $\mu=\wh\mu$, scale with $|\ep_0|^2$,
and the ones for $\mu\ne\wh\mu$ with  $|\ep_0|$, cf.,  
Eq.~\bref{f5} and Eq.~\bref{f6}.

\end{document}